\documentclass[final]{clear2026} 

\newcommand{\sE}[0]{\mathscr{E}}
\newcommand{\sF}[0]{\mathscr{F}}
\newcommand{\sH}[0]{\mathscr{H}}
\newcommand{\sG}[0]{\mathscr{G}}
\newcommand{\sP}[0]{\mathscr{P}}

\newcommand{\EE}[0]{\mathbb{E}}
\newcommand{\KK}[0]{\mathbb{K}}

\newcommand{\NN}[0]{\mathbb{N}}
\newcommand{\PP}[0]{\mathbb{P}}
\newcommand{\QQ}[0]{\mathbb{Q}}
\newcommand{\RR}[0]{\mathbb{R}}

\newcommand{\cC}[0]{\mathcal{C}}

\newcommand{\cS}[0]{\mathcal{S}}

\newcommand{\cV}[0]{\mathcal{V}}

\newcommand{\bX}[0]{\mathbf{X}}

\newcommand{\bp}[0]{\boldsymbol{p}}

\newcommand{\bx}[0]{\boldsymbol{x}}

\newcommand{\iPP}[2]{\PP^{\textnormal{do}(#1,#2)}}
\newcommand{\iEE}[2]{\EE^{\textnormal{do}(#1,#2)}}
\newcommand{\iKK}[2]{\KK^{\textnormal{do}(#1,#2)}}
\newcommand{\iK}[3]{K^{\textnormal{do}(#1,#2)}_{#3}}
\newcommand{\iS}[3]{\cS^{\textnormal{do}(#1,#2)}_{#3}}

\newcommand{\ind}[0]{\mathbf{1}}
\DeclareMathOperator*{\argmax}{arg\,max}

\newcommand{\IndepAux}[2]{
  \ifx#1\displaystyle \def\k{-10mu}\else
  \ifx#1\textstyle    \def\k{-9mu}\else
  \ifx#1\scriptstyle  \def\k{-7mu}\else
                       \def\k{-6mu}\fi\fi\fi
  \nonscript\mskip\thickmuskip\nobreak
  \mathord{\perp\mkern\k\perp}_{\!#2}%
  \nobreak\nonscript\mskip\thickmuskip
}

\newcommand{\ins}[0]{\textnormal{Ins}}
\newcommand{\pay}[0]{\textnormal{Pay}}
\newcommand{\dan}[0]{\textnormal{Dan}}



\makeatletter
\NewDocumentEnvironment{savetheorem}{m +b}{
  \begin{theorem}\label{#1}#2\end{theorem}%
  \expandafter\gdef\csname rest@body@#1\endcsname{#2}%
}{}%
\newcommand{\printtheorem}[1]{%
  \par\noindent\textbf{Theorem~\ref{#1}. }\itshape
  \csname rest@body@#1\endcsname\par\normalfont
}

\NewDocumentEnvironment{saveprop}{m +b}{%
  \begin{proposition}\label{#1}#2\end{proposition}%
  \expandafter\gdef\csname rest@body@#1\endcsname{#2}%
}{}%
\newcommand{\printprop}[1]{%
  \par\noindent\textbf{\propositionref{#1}. }\itshape
  \csname rest@body@#1\endcsname\par\normalfont
}
\makeatother

\NewDocumentEnvironment{savelemma}{m +b}{%
  \begin{lemma}\label{#1}#2\end{lemma}%
  \expandafter\gdef\csname rest@body@#1\endcsname{#2}%
}{}%
\newcommand{\printlemma}[1]{%
  \par\noindent\textbf{\lemmaref{#1}. }\itshape
  \csname rest@body@#1\endcsname\par\normalfont
}
\makeatother

\usepackage{amsfonts}
\usepackage{mathtools}
\usepackage{thmtools}

\usepackage{xparse}
\usepackage{fix-cm}
\usepackage{booktabs}
\usepackage{hyperref}
\usepackage{times}
\usepackage{lmodern}
\usepackage[mathscr]{eucal}
\usepackage{dirtytalk}
\usepackage[shortlabels]{enumitem}
\usepackage[T1]{fontenc}
\usepackage{tikz}
\usetikzlibrary{shapes.geometric,arrows,positioning}
\usepackage{multirow}
\usepackage{footmisc}
\usepackage{wrapfig}
\allowdisplaybreaks
\usepackage{refcount}

\title[Causal effects in causal spaces]{A fine-grained look at causal effects in causal spaces}

\clearauthor{\Name{Junhyung Park} \Email{jun.park@inf.ethz.ch}\AND
 \Name{Yuqing Zhou} \Email{yuqzhou@student.ethz.ch}\\
 \addr ETH Z\"urich, Switzerland}

\begin{document}

\maketitle

\begin{abstract}%
  The notion of causal effect is fundamental across many scientific disciplines. Traditionally, quantitative researchers have studied causal effects at the level of variables; for example, how a certain drug dose (\(W\)) causally affects a patient's blood pressure (\(Y\)). However, in many modern data domains, the raw variables---such as pixels in an image or tokens in a language model---do not have the semantic structure needed to formulate meaningful causal questions. In this paper, we offer a more fine-grained perspective by studying causal effects at the level of \emph{events}, drawing inspiration from probability theory, where core notions such as independence are first given for events and \(\sigma\)-algebras, before random variables enter the picture. Within the measure-theoretic framework of causal spaces, a recently introduced axiomatisation of causality, we first introduce several binary definitions that determine whether a causal effect is present, as well as proving some properties of them linking causal effect to (in)dependence under an intervention measure. Further, we provide quantifying measures that capture the strength and nature of causal effects on events, and show that we can recover the common measures of treatment effect as special cases. %
\end{abstract}


\section{Introduction}\label{sec:introduction}
The notion of \emph{cause}, or \emph{causal effect}, is central to the human psychology, as well as numerous academic domains. It is deeply engrained in everyday language via constructions such as \say{if}, \say{so} or \say{because} \citep{kahneman2011thinking}. It also forms the backbone of virtually all scientific enquiries; examples of causal statements include \say{gravity causes apples to fall} \citep{cohen1981newton}, \say{\(\text{CO}_2\) emission is causing climate change} \citep{lee2023ipcc} and \say{vaccination reduces the risk of fatality from Covid-19} \citep{ledford2020covid}. The goal of this paper is to give formal mathematical definitions of causal effects, in many nuanced forms. 

Causality is rarely deterministic. When a government introduces a policy, it may have rough ideas or hopes about its impact, but it cannot know exactly what will occur as a result; similarly when a company runs an advertising campaign, when a doctor prescribes a drug, or when a user prompts a language model. Thus, naturally, a crucial aspect of causality is \emph{randomness}. The latter concept is axiomatised by \emph{probability spaces} \citep{kolmogorov1933foundations}, grounded in measure theory and forming the backbone of \emph{probability theory}. However, probability spaces do not, by themselves, encode any causality. Recently, an analogous axiomatic framework of causality called \emph{causal spaces} was introduced \citep{park2023measure}. It was shown that causal spaces generalise and unify popular frameworks of causality, such as the \emph{structural causal models} (SCMs) \citep{pearl2009causality} and the \emph{potential outcomes} \citep{imbens2015causal}. We use causal spaces as the basis for our definitions of causal effects. 

Existing approaches (cf. \sectionref{subsec:treatment_effect} \& \appendixref{sec:related}) define causal effects at the level of \emph{variables}, tacitly assumed to be semantically meaningful (e.g., \say{salary} following a policy, \say{blood pressure} after a medical intervention). As a key distinction, we define causal effects first on \emph{events}, then extend to \(\sigma\)-algebras. In many modern data domains, the raw variables, such as pixels in images or tokens in language models, often lack the right semantics for meaningful causal statements. However, we can define events such as \say{there is a cat in this image}, \say{this image is in greyscale}, \say{this text is about politics} or \say{this text is in Chinese} to which an image or a piece of text either belongs or does not belong, based on which causal effects can be analysed in a semantically meaningful way. Further, the extension to \(\sigma\)-algebras facilitates graded, rather than binary, encoding of outcome characteristics. 

The paper is broadly organised in two parts. 
\begin{itemize}
    \item In \sectionref{sec:binary}, we give binary definitions of causal effects, as well as their conditional and post-intervention variants. The definitions specify when an outcome or an event has a causal effect on an event or a \(\sigma\)-algebra. 
    \item In \sectionref{sec:quantifying_measures}, we propose ways of capturing the nature and strength of causal effects. 
\end{itemize}
Our work has parallels with how the notion of (conditional) independence develops in probability theory. A binary notion of independence is first given for events, via \(\PP(A\cap B)=\PP(A)\PP(B)\), and then for \(\sigma\)-algebras. Following these, many quantities are proposed to capture the nature and strength of (conditional) dependence---see \appendixref{subsec:independence}.

In this paper, we do not consider identification, hypothesis tests of causal effects, or estimating the quantifying measures, from finite samples: our focus is solely on population-level quantities. 

\subsection{Related works}

\paragraph{Treatment effect}\label{subsec:treatment_effect}
We recall the \emph{average treatment effect} (ATE) \citep{rubin1974estimating} and its many variants. These are the most widely used definitions of causal effects, and are most often given in the \emph{potential outcomes framework} \citep{imbens2015causal,hernan2020what}.\footnote{It can also be done in the \emph{structural causal models} (SCM) framework \citep[p.112]{peters2017elements}.} Let \(W\in\{0,1\}\) be a binary treatment variable, and \(Y_0\) and \(Y_1\) be potential outcome variables corresponding to control (\(W=0\)) and treatment (\(W=1\)) respectively. Then the ATE is
\[\text{ATE}=\EE[Y_1-Y_0].\]
The \emph{conditional} average treatment effect (CATE) \citep{rosenbaum1983central} was proposed to take \emph{treatment effect heterogeneity} into account, i.e., the effect of treatment can differ across sub-populations. It is defined with respect to a conditioning set of variables \(\bX\):
\[\text{CATE}(\bx)=\EE[Y_1-Y_0\mid\bX=\bx].\]
Many variants have been studied, including, but not limited to, the distributional treatment effect (DTE) \citep{heckman1997making}, conditional distributional treatment effect (CDTE) \citep{park2021conditional,kallus2023robust}, individual treatment effect (ITE) \citep{rubin1974estimating}, local average treatment effect (LATE) \citep{imbens1994identification}, average treatment effect on the treated (ATT) \citep{heckman1985alternative}, quantile treatment effect (QTE) \citep{firpo2007efficient} and marginal average treatment effect (MATE) \citep{heckman2005structural}. As mentioned earlier, common to these definitions is that causal effects are defined for \emph{variables} rather than \emph{events}. They also \emph{quantify} magnitudes of effects from the outset, rather than first providing binary, \say{effect/no effect} notions. We review their precise definitions, as well as some other discussions on related works, in \appendixref{sec:related}. 

\paragraph{Causal abstraction}\label{subsec:causal_abstraction}
A complementary line of work is causal abstraction \citep{beckers2019abstracting,buchholz2024products}, which likewise aims to study causal effects at a semantic level rather than at the level of \say{raw} variables. There is a particularly high level of interest in this technique in the context of \emph{mechanistic interpretability} \citep{geiger2025causal}. In that literature, one typically defines an explicit abstraction map (or representation) \(\alpha\) from a low-level model/state space to a higher-level set of abstract variables, and then defines and compares causal effects in the abstracted space. In this way, causal abstraction can talk about the causal influence of variables that are not the original coordinates (pixels, tokens, etc.), but instead live in a chosen or learned representational space. Our approach shares the same motivation---capturing causality at the right semantic granularity---but differs in what must be postulated: we do not require selecting an abstraction function \(\alpha\) or constructing a separate abstract causal model. Instead, we define causal effects directly in the original causal space on events and on \(\sigma\)-algebras, so that \say{high-level} questions can be posed as events (or collections of events) without leaving the underlying space or committing to a particular representation map.

\subsection{Preliminaries \& notations}\label{subsec:preliminaries}
Let \((\Omega,\sH,\PP)\) be a probability space. For each event \(A\in\sH\), we write \(\ind_A:\Omega\to\{0,1\}\) for the indicator, and for each outcome \(\omega\in\Omega\), we write \(\delta_\omega:\sH\to\{0,1\}\) for the delta measure. 

We use the somewhat uncommon notation of conditional measures from \citet{cinlar2011probability}. The conditional measure of \(A\) given a \(\sigma\)-algebra \(\sG\) is the \(\sG\)-measurable random variable \(\PP_\sG(\cdot,A)\). Similarly, for an event \(G\in\sH\), we denote the conditional measure of \(A\) given \(G\) by \(\PP_G(A)\), which is \(\frac{\PP(G\cap A)}{\PP(G)}\) if \(\PP(G)>0\) and undefined if \(\PP(G)=0\). Note that \(\PP_\sG(\cdot,A)\) and \(\PP_G(A)\) are different objects---the former is a random variable, and the latter a number. 

We now formally introduce the causal space framework \citep{park2023measure}. We require that the measurable space be in product form. Let \(T\) be the index set of the product. Then taking, for each \(t\in T\), a measurable space \((\Omega_t,\sE_t)\), we have the product measurable space
\begin{equation*}
    (\Omega,\sH)=\otimes_{t\in T}(\Omega_t,\sE_t)=(\times_{t\in T}\Omega_t,\otimes_{t\in T}\sE_t).
\end{equation*}
For \(S\in\sP(T)\), where \(\sP(T)\) denotes the power set of \(T\), we write \(\sH_S\) for the sub-\(\sigma\)-algebra of \(\sH\) generated by measurable rectangles \(\times_{t\in T}A_t\), where \(A_t\in\sE_t\) differs from \(\Omega_t\) only for finitely many \(t\in S\). We also write \(\Omega_S=\times_{s\in S}\Omega_s\), and for each \(\omega=(\omega_t)_{t\in T}\), we write \(\omega_S=(\omega_s)_{s\in S}\in\Omega_S\), where \(\omega_s\in\Omega_s\). 

A causal space \citep[Definition 2.2]{park2023measure} is defined as follows. 
\begin{definition}[Causal space]\label{def:causal_space}
    A \emph{causal space} is the quadruple \((\Omega,\sH,\PP,\KK)\), where \((\Omega,\sH)\) is a measurable space with the above product structure, \(\PP\) is a probability measure on \((\Omega,\sH)\) and \(\KK=\{K_S:S\subseteq T\}\), called the \emph{causal mechanism}, is a collection of transition probability kernels \(K_S\) from \((\Omega,\sH_S)\) into \((\Omega,\sH)\), called the \emph{causal kernel on \(\sH_S\)}. This means that each \(K_S\) is a function \(K_S:\Omega\times\sH\to[0,1]\), such that
    \begin{itemize}
        \item for each \(\omega\in\Omega\), the function \(B\mapsto K_S(\omega,B):\sH\to[0,1]\) is a probability measure, and
        \item for each \(B\in\sH\), the function \(\omega\mapsto K_S(\omega,B):\Omega\to[0,1]\) is \(\sH_S\)-measurable.
    \end{itemize}
    The causal kernels must satisfy the following axioms:
    \begin{enumerate}[(i)]
        \item for all \(A\in\sH\) and \(\omega\in\Omega\), we have \(K_\emptyset(\omega,A)=\PP(A)\);
        \item for all \(\omega\in\Omega\), \(A\in\sH\) and \(B\in\sH_S\), we have \(K_S(\omega,A\cap B)=\ind_B(\omega)K_S(\omega,A)\).
    \end{enumerate}
\end{definition}
The probability measure \(\PP\) is the \emph{observational measure}, and \(\KK\) encodes the causal information. Intuitively, for each \(S\subseteq T\) and \(\omega\in\Omega\), the measure \(K_S(\omega,\cdot)\) describes the distribution we get on the whole space after we forcing the coordinates indexed by \(S\) to take the value \(\omega_S\). Axioms (i) and (ii) formalise this intuition by recovering the observational measure in the absence of intervention and enforcing consistency with the imposed value on the intervened coordinates. This is analogous to the interventional distribution in an SCM, although the two frameworks encode causality differently: SCMs do so through structural equations, whereas causal spaces take \(\{K_S\}_{S\subseteq T}\) as primitive objectives. We now introduce the corresponding notion of \emph{interventions} \citep[Definition 2.3]{park2023measure}:
\begin{definition}[Intervention]\label{def:interventions}
    Let us take a causal space \((\Omega,\sH,\PP,\KK)\), an intervention set \(U\subseteq T\) and a measure \(\QQ\) on \((\Omega,\sH_U)\). An \emph{intervention on \(\sH_U\) via \(\QQ\)} yields a new causal space \((\Omega,\sH,\iPP{U}{\QQ},\iKK{U}{\QQ})\), where the \emph{intervention measure} \(\iPP{U}{\QQ}\) is a probability measure on \((\Omega,\sH)\) defined, for \(A\in\sH\), by
    \[\iPP{U}{\QQ}(A)=\int\QQ(d\omega)K_U(\omega,A)\]
    and \(\iKK{U}{\QQ}=\{\iK{U}{\QQ}{S}:S\subseteq T\}\) is the \emph{intervention causal mechanism}, where
    \[\iK{U}{\QQ}{S}(\omega_S,A)=\int\QQ(d\omega'_{U\setminus S})K_{S\cup U}((\omega_S,\omega'_{U\setminus S}),A).\]
\end{definition}
Hence, causal kernels of the original causal space encode what the new measure and causal kernels will be after an intervention. We denote by \(\iEE{U}{\QQ}\) the expectation with \(\iPP{U}{\QQ}\). 

\subsection{Running example}\label{subsec:example}
To illustrate our theory, we give a simple running example to accompany the definitions throughout the paper. In \appendixref{sec:lm}, we present another example with a language model. 
\begin{example}[Travel insurance]\label{ex:insurance}
    Suppose that we want to model how much money a traveller has to pay towards accident-related costs, depending on the danger level of the trip and whether or not the traveller takes out insurance. The insurance costs 30 CHF, and if an accident happens, the traveller has to pay 1000 CHF. We take the outcome sets
    \[\Omega_\dan=\{N,L,H\},\quad\Omega_\ins=\{Y,N\},\quad\Omega_\pay=\{0,30,1000\},\]
    where \(Y\) and \(N\) stands for Yes/No, \(N\), \(L\) and \(H\) stands for No/Low/High danger. The observational measure \(\PP\) is given in \tableref{tab:insurance_observational}, and the causal kernels \(K_\ins(Y,\cdot)\) and \(K_\ins(N,\cdot)\) corresponding to making the traveller buy or not buy insurance are given in \tableref{tab:insurance_yes,tab:insurance_no}. 
    \begin{table}[t]
        \centering
        \begin{tabular}{r|cccccc|c}
            &\multicolumn{6}{c}{\((\omega_\dan,\omega_\ins)\)}\\
            \(\PP\)&\((N,Y)\)&\((N,N)\)&\((L,Y)\)&\((L,N)\)&\((H,Y)\)&\((H,N)\)&Sum\\\hline
            \(0\)&\(0\)&\(0.09\)&\(0\)&\(0.345\)&\(0\)&\(0.04875\)&\(0.48375\)\\
            \(\omega_\pay\qquad30\)&\(0.01\)&\(0\)&\(0.35\)&\(0\)&\(0.15\)&\(0\)&\(0.51\)\\
            \(1000\)&\(0\)&\(0\)&\(0\)&\(0.005\)&\(0\)&\(0.00125\)&\(0.00625\)\\\hline
            Sum&\(0.01\)&\(0.09\)&\(0.35\)&\(0.35\)&\(0.15\)&\(0.05\)&\(1\)
        \end{tabular}
        \caption{The observational measure \(\PP\) on the danger of the trip, whether or not the traveller gets insurance and how much the traveller pays towards accident-related costs.}
        \label{tab:insurance_observational}
    \end{table}
    \begin{table}[t]
        \centering
        \begin{tabular}{r|cccccc|c}
            &\multicolumn{6}{c}{\((\omega_\dan,\omega_\ins)\)}\\
            \(K_\ins(Y,\cdot)\)&\((N,Y)\)&\((N,N)\)&\((L,Y)\)&\((L,N)\)&\((H,Y)\)&\((H,N)\)&Sum\\\hline
            \(0\)&\(0\)&\(0\)&\(0\)&\(0\)&\(0\)&\(0\)&\(0\)\\
            \(\omega_\pay\qquad30\)&\(0.05\)&\(0\)&\(0.65\)&\(0\)&\(0.3\)&\(0\)&\(1\)\\
            \(1000\)&\(0\)&\(0\)&\(0\)&\(0\)&\(0\)&\(0\)&\(0\)\\\hline
            Sum&\(0.05\)&\(0\)&\(0.65\)&\(0\)&\(0.3\)&\(0\)&\(1\)
        \end{tabular}
        \caption{The causal kernel \(K_\ins(Y,\cdot)\) corresponding to making the traveller buy insurance.}
        \label{tab:insurance_yes}
    \end{table}
    \begin{table}[t]
        \centering
        \begin{tabular}{r|cccccc|c}
            &\multicolumn{6}{c}{\((\omega_\dan,\omega_\ins)\)}\\
            \(K_\ins(N,\cdot)\)&\((N,Y)\)&\((N,N)\)&\((L,Y)\)&\((L,N)\)&\((H,Y)\)&\((H,N)\)&Sum\\\hline
            \(0\)&\(0\)&\(0.3\)&\(0\)&\(0.59\)&\(0\)&\(0.095\)&\(0.985\)\\
            \(\omega_\pay\qquad30\)&\(0\)&\(0\)&\(0\)&\(0\)&\(0\)&\(0\)&\(0\)\\
            \(1000\)&\(0\)&\(0\)&\(0\)&\(0.01\)&\(0\)&\(0.005\)&\(0.015\)\\\hline
            Sum&\(0\)&\(0.3\)&\(0\)&\(0.6\)&\(0\)&\(0.1\)&\(1\)
        \end{tabular}
        \caption{The causal kernel \(K_\ins(N,\cdot)\) corresponding to making the traveller uninsured.}
        \label{tab:insurance_no}
    \end{table}

    \tableref{tab:insurance_yes} indicates that, after an intervention to make the traveller buy insurance, the payment is always 30 CHF and the insurance status is always "Yes". The probabilities of the traveller choosing the danger level to be "No", "Low" and "High" become 0.05, 0.65 and 0.3 respectively. Since the corresponding observational probabilities in \tableref{tab:insurance_observational} are 0.10, 0.70 and 0.20, the intervention makes higher danger levels slightly more likely.

    \tableref{tab:insurance_no} indicates that, after an intervention forcing the traveller not to buy insurance, the payment is 0 CHF with probability 0.985 and 1000 CHF with probability 0.015, while the payment 30 CHF occurs with probability 0. Likewise, the insurance status is now always "No". The probabilities of the danger levels to be "No", "Low" and "High" are 0.30, 0.60 and 0.10 respectively. Compared with \tableref{tab:insurance_observational}, this indicates that forcing the traveller not to buy insurance slightly reduces the danger levels.
\end{example}

\section{Binary definitions}\label{sec:binary}
In this section, we give binary, yes/no definitions of causal effects. In \sectionref{subsec:causal_effect}, we give the vanilla definition of causal effects, of whether the marginal observational measure \(\PP\) changes as a result of an intervention. In \sectionref{subsec:conditional_causal_effect}, we define causal effects conditioned either on an event \(G\in\sH\) or a sub-\(\sigma\)-algebra \(\sG\subseteq\sH\), i.e., whether the conditional measures \(\PP_G\) and \(\PP_\sG\) change. In \sectionref{subsec:post-intervention}, we define post-intervention causal effects, which characterises the effect of an additional intervention after a first one has already been applied.

Due to space constraints, we only focus on \emph{active} causal effects and their negations in the main body, and defer the full trichotomy into no, active and dormant effects to \appendixref{sec:no_dormant}. 

\subsection{Causal effect}\label{subsec:causal_effect}
The following is the most basic definition of (active) causal effects. 
\begin{definition}[Active causal effect]\label{def:active_causal_effect}
    Let us take a causal space \((\Omega,\sH,\PP,\KK)\), an intervention set \(U\subseteq T\), an outcome \(\omega\in\Omega\) and events \(A,B\in\sH\). 
    \begin{enumerate}[(i)]
        \item If \(K_U(\omega,A)=\PP(A)\), then we say that \(\omega\) has \emph{no active \(U\)-causal effect} on \(A\). 
        \item If \(K_U(\omega,A)=\PP(A)\) for all \(\omega\in B\), then we say that \(B\) has \emph{no active \(U\)-causal effect} on \(A\).
    \end{enumerate}
    For a sub-\(\sigma\)-algebra \(\sF\) of \(\sH\), we say that \(\omega\) (resp.\ \(B\)) has no active \(U\)-causal effect on \(\sF\) if \(\omega\) (resp.\ \(B\)) has no active \(U\)-causal effect on any \(A\in\sF\). 
\end{definition}
When the intervention set \(U\) is clear from context, we also simply say that \(\omega_U\) (or \(\omega\)) has an active causal effect on \(A\).

\citet[Definition B.2]{park2023measure} gave a coarser version of the above, defining \(\sH_U\) to have no active causal effect on \(A\) if \(K_U(\omega,A)=\PP(A)\) for all \(\omega\in\Omega\), coinciding with \definitionref{def:active_causal_effect}(ii) when \(B=\Omega\). \definitionref{def:active_causal_effect} allows a more fine-grained analysis of causal effects---the quantities having active causal effects are individual outcomes and events, rather than entire \(\sigma\)-algebras (or random variables). This is necessary in many modern data domains; for example, in a language model, instead of saying that \say{the prompt has a causal effect on the output text being about sports} (which is trivially true), we want to identify specific (sets of) prompts that have this causal effect. 

Note that, depending on the intervention set, it is possible for an outcome \(\omega\in\Omega\) to have an active \(U\)-causal effect on an event \(A\), yet for the same outcome to have no active \(V\)-causal effect on the same event \(A\). Note also that there is no \emph{control} level, unlike the treatment effects defined in \sectionref{subsec:treatment_effect}. Instead, the causal kernels \(K_U(\omega,\cdot)\) are compared against the observational measure \(\PP(\cdot)\). 


In the following result, we show that no active causal effect means independence under the intervention measure (see \definitionref{def:interventions}). The proof is in \appendixref{sec:proofs}. 
\begin{savelemma}{lem:independence}
    Let us take a causal space \((\Omega,\sH,\PP,\KK)\), an intervention set \(U\subseteq T\) and an event \(A\in\sH\). If \(\Omega\) has no active \(U\)-causal effect on \(A\), then with respect to the intervention measure \(\iPP{U}{\QQ}\) with any measure \(\QQ\) on \((\Omega,\sH_U)\), the event \(A\) and the \(\sigma\)-algebra \(\sH_U\) are independent.
\end{savelemma}
Let us now return to our running example to illustrate \definitionref{def:active_causal_effect}, by showing that buying insurance has a causal effect on the probability of the traveller paying 1000 CHF. 
\begin{example}
    Write \(A=\{\omega_\pay=1000\}\) for the event that the traveller pays \(1000\) CHF. From \tableref{tab:insurance_observational,tab:insurance_yes,tab:insurance_no}, the marginal probabilities of \(A\) under the observational state and after interventions to force the traveller to buy/not buy insurance are respectively
    \[\PP(A)=0.00625,\qquad K_\ins(Y,A)=0\neq\PP(A),\qquad K_\ins(N,A)=0.015\neq\PP(A).\]
    Hence, both outcomes \(\omega_\ins=Y\) and \(\omega_\ins=N\) have active causal effects on the event \(A\). 
\end{example}

\subsection{Conditional causal effect}\label{subsec:conditional_causal_effect}
Conditional causal effect is an important concept, as causal effects often differ across subpopulations (\emph{treatment effect heterogeneity}) \citep{heckman1997making}. For example, the injection of the prompt \say{Answer in a politically neutral tone} in a language model will have a marked causal effect on the output if earlier parts of the prompt were on politics, but little to no effect otherwise. In the classical definitions of conditional treatment effects (e.g., CATE or CDTE, cf. \sectionref{subsec:treatment_effect} \& \appendixref{sec:related}), one conditions on a \emph{variable}, which is again assumed to be semantically meaningful. In this section, we introduce binary definitions of conditional causal effects, conditioning on general \emph{events} \(G\) and \emph{\(\sigma\)-algebras} \(\sG\). 

In this section, we write \(\iPP{U}{\delta_\omega}(\cdot)=K_U(\omega,\cdot)\) interchangeably, the former preferred when we want to condition on this measure, to avoid having two subscripts. It is trivial from the definition of interventions (\definitionref{def:interventions}) that these two measures are equal. We first present the simpler definition of (active) causal effects conditioned on an event \(G\). 
\begin{definition}[Conditional active causal effect I]\label{def:active_causal_effect_conditional_event}
    Take a causal space \((\Omega,\sH,\PP,\KK)\), an intervention set \(U\subseteq T\), an outcome \(\omega\in\Omega\), events \(A,B,G\in\sH\) and a \(\sigma\)-algebra \(\sF\). 
    \begin{enumerate}[(i)]
        \item If \(\PP(G)>0\), \(\iPP{U}{\delta_\omega}(G)>0\) and \(\iPP{U}{\delta_\omega}_G(A)=\PP_G(A)\), then we say that \(\omega\) has \emph{no active \(U\)-causal effect} on \(A\) \emph{conditioned on} \(G\). 
        \item If no \(\omega\in B\) has an active \(U\)-causal effect on \(A\) conditioned on \(G\), then we say that \(B\) has no active \(U\)-causal effect on \(A\) conditioned on \(G\).  
    \end{enumerate}
    We say that \(\omega\) (resp.\ \(B\)) has no active \(U\)-causal effect on \(\sF\) conditioned on \(G\) if \(\omega\) (resp.\ \(B\)) has no active \(U\)-causal effect on any \(A\in\sF\) conditioned on \(G\). 
\end{definition}
The slight complication in \definitionref{def:active_causal_effect_conditional_event} is due to the fact that conditional probabilities given \(G\) are only well-defined if \(G\) has positive measure. If we condition on the full event \(\Omega\), then we recover \definitionref{def:active_causal_effect}. Otherwise, we require \(G\) to have positive measure under both \(\PP\) and \(\iPP{U}{\delta_\omega}\), before we compare \(\PP_G(A)\) and \(\iPP{U}{\delta_\omega}_G(A)\). Just as conditional independence cannot be determined if the conditioning event has zero measure, the existence of conditional active causal effect given \(G\) is undefined if \(G\) has zero measure under either \(\PP\) or \(\iPP{U}{\delta_\omega}\). 

The more involved case is when we condition on a \(\sigma\)-algebra \(\sG\), rather than an event \(G\). We need to be careful when conditioning an interventional measure \(\iPP{U}{\delta_\omega}\) on a \(\sigma\)-algebra \(\sG\), because there are two outcomes in play: one for the conditioning and another for the intervention. Explicitly, given an event \(A\in\sH\), using \(\omega\) for intervention and \(\tilde{\omega}\) for conditioning, we have the function
\[(\omega,\tilde{\omega})\mapsto\iPP{U}{\delta_\omega}_\sG(\tilde{\omega},A).\]
For a fixed \(\omega\), the map \(\tilde{\omega}\mapsto\iPP{U}{\delta_\omega}_\sG(\tilde{\omega},A)\) is \(\sG\)-measurable, and is defined \(\iPP{U}{\delta_\omega}\)-almost surely. With this in mind, we define conditional (active) causal effect given \(\sG\). 
\begin{definition}[Conditional active causal effect II]\label{def:active_causal_effect_conditional}
    Take a causal space \((\Omega,\sH,\PP,\KK)\), an intervention set \(U\subseteq T\), an outcome \(\omega\in\Omega\), events \(A,B\in\sH\) and sub-\(\sigma\)-algebras \(\sG,\sF\subseteq\sH\). 
    \begin{enumerate}[(i)]
        \item Suppose that \(\PP\) and \(\iPP{U}{\delta_\omega}\) are mutually absolutely continuous on \(\sG\). We say that \(\omega\) has no \emph{active \(U\)-causal effect} on \(A\) conditioned on \(\sG\) if \(\PP_\sG(\tilde{\omega},A)=\iPP{U}{\delta_\omega}_\sG(\tilde{\omega},A)\) for \(\PP\)- (or equivalently, \(\iPP{U}{\delta_\omega}\)-)almost all \(\tilde{\omega}\in\Omega\).
        \item If no \(\omega\in B\) has an active \(U\)-causal effect on \(A\) conditioned on \(\sG\), then we say that \(B\) has no active \(U\)-causal effect on \(A\) conditioned on \(\sG\). 
    \end{enumerate}
    We say that \(\omega\) (resp.\ \(B\)) has no active \(U\)-causal effect on \(\sF\) conditioned on \(\sG\) if \(\omega\) (resp.\ \(B\)) has no active \(U\)-causal effect on any \(A\in\sF\) conditioned on \(\sG\). 
\end{definition}
The complication in \definitionref{def:active_causal_effect_conditional} arises from the fact that conditional measures given \(\sG\) are only well defined almost surely. However, in essence, \definitionref{def:active_causal_effect_conditional} is again just a direct conditional analogue of \definitionref{def:active_causal_effect}; indeed, if the conditioning \(\sigma\)-algebra is trivial, i.e., \(\sG=\{\emptyset,\Omega\}\), then we recover \definitionref{def:active_causal_effect}. When \(\sG\) is not trivial, \(\PP\) and \(\iPP{U}{\delta_\omega}\) are required to be mutually absolutely continuous on \(\sG\), before comparing \(\PP_\sG(\cdot,A)\) and \(\iPP{U}{\delta_\omega}_\sG(\cdot,A)\). If this were not the case, then in regions with zero measure under one measure and positive measure under the other, the existence of an active causal effect cannot be determined. Note that mutual absolute continuity is only required to hold on \(\sG\), not on the full \(\sigma\)-algebra \(\sH\). 


The following result is a conditional analogue of \lemmaref{lem:independence}: it shows that no active conditional causal effect means conditional independence under the intervention measure. The proof requires some gymnastics with monotone convergence theorem, and is in \appendixref{sec:proofs}. 
\begin{saveprop}{prop:conditional_independence}
    Take a causal space \((\Omega,\sH,\PP,\KK)\), an intervention set \(U\subseteq T\), events \(A,G\in\sH\) and a sub-\(\sigma\)-algebra \(\sG\subseteq\sH\). If \(\Omega\) has no active \(U\)-causal effect on \(A\) conditioned on \(G\) (resp.\ \(\sG\)), then with respect to the intervention measure \(\iPP{U}{\QQ}\) with any \(\QQ\) on \((\Omega,\sH_U)\), the event \(A\) and the \(\sigma\)-algebra \(\sH_U\) are conditionally independent given \(G\) (resp.\ \(\sG\)).
\end{saveprop}
Let us return to the running example again. According to \definitionref{def:active_causal_effect_conditional_event}, we show that, conditioned on the event that the trip has no danger, buying insurance has no causal effect on the probability of the traveller paying 1000 CHF. However, given that the trip is highly dangerous, buying insurance does have a causal effect on the same event. 
\begin{example}\label{ex:insurance_conditional_effect}
    Recall that we had \(A=\{\omega_\pay=1000\}\) for the event that the traveller pays \(1000\) CHF. Write \(G=\{\omega_\dan=N\}\) and \(H=\{\omega_\dan=H\}\) for the events that the danger level of the trip is non-existent and high. From the first two columns of \tableref{tab:insurance_observational,tab:insurance_yes}, the conditional probabilities of \(A\) the observational measure and after interventions to force the traveller to buy insurance given that the trip has no danger are
    \[\PP_G(A)=0,\qquad\iPP{\ins}{Y}_G(A)=0.\]
    This means that, conditioned on \(G\), \(\omega_\ins=Y\) has no causal effect on \(A\), unlike the marginal case. However, conditioned on \(H\), the last two columns of \tableref{tab:insurance_observational,tab:insurance_yes} tell us 
    \[\PP_H(A)=\frac{0.00125}{0.2}=0.00625,\quad\iPP{\ins}{Y}_H(A)=0\neq\PP_H(A),\]
    so \(\omega_\ins=Y\) does have a causal effect on \(A\) conditioned on \(H\). 
\end{example}

\subsection{Post-intervention causal effect}\label{subsec:post-intervention}
As a final variant of (active) causal effects, we present the definition of \emph{post-intervention} causal effects. This definition identifies cases in which, once an intervention takes place on \(\sH_V\), no further intervention on \(\sH_U\) has an effect on \(A\). This can be due to \(\sH_V\) being the \emph{mediator} of the causal effect of \(\sH_U\) on \(A\), or due to an intervention on \(\sH_V\) creating a different environment in which interventions on \(\sH_U\) have no active causal effect on \(A\). 
\begin{definition}[Post-intervention causal effect]\label{def:active_causal_effect_post-intervention}
    Let us take a causal space \((\Omega,\sH,\PP,\KK)\), subsets \(U,V\subseteq T\), an outcome \(\omega\in\Omega\) and events \(A,B\in\sH\). 
    \begin{enumerate}[(i)]
        \item If \(K_{U\cup V}((\omega_U,\omega'_{V\setminus U}),A)=K_V((\omega_{U\cap V},\omega'_{V\setminus U}),A)\) for all \(\omega'_{V\setminus U}\in\Omega_{V\setminus U}\), then we say that \(\omega\) has no active \(U\)-causal effect on \(A\) after intervening on \(\sH_V\). 
        \item If no \(\omega\in B\) has an active \(U\)-causal effect on \(A\) after intervening on \(\sH_V\), then we say that \(B\) has no active \(U\)-causal effect on \(A\) after intervening on \(\sH_V\). 
    \end{enumerate}
    For a sub-\(\sigma\)-algebra \(\sF\) of \(\sH\), we say that \(\omega\) (resp.\ \(B\)) has no active \(U\)-causal effect on \(\sF\) after intervening on \(\sH_V\) if \(\omega\) (resp.\ \(B\)) has no active \(U\)-causal effect on any \(A\in\sF\) after intervening on \(\sH_V\). 
\end{definition}
Note that this is different to conditional causal effect---indeed, that conditioning and intervening are different is the whole premise of any theory of causality. The intuition behind this definition becomes clearer when we give the interventional causal kernels \(\iK{V}{\QQ}{U}\) explicitly in the following result, which is proved in \appendixref{sec:proofs}. 
\begin{saveprop}{prop:post-intervention}
    Let us take a causal space \((\Omega,\sH,\PP,\KK)\), subsets \(U,V\subseteq T\) and an event \(A\in\sH\). Suppose that \(\omega\) has no active \(U\)-causal effect on \(A\) after intervening on \(\sH_V\). 
    \begin{enumerate}[(i)]
        \item If \(U\cap V=\emptyset\), then in the causal space \((\Omega,\sH,\iPP{V}{\QQ},\iKK{V}{\QQ})\) obtained via an intervention on \(\sH_V\) with any measure \(\QQ\) on \((\Omega,\sH_V)\), we have that \(\omega\) has no active \(U\)-causal effect on \(A\), i.e., \(\iK{V}{\QQ}{U}(\omega,A)=\iPP{V}{\QQ}(A)\). 
        \item In the causal space \((\Omega,\sH,\iPP{U}{\QQ},\iKK{U}{\QQ})\) obtained after an intervention on \(\sH_U\) with any \(\QQ\), we have that \(\omega\) continues to have no active \(U\)-causal effect on \(A\) after intervening on \(\sH_V\). 
    \end{enumerate}
\end{saveprop}
We can see in part (i) that, indeed, once we intervene on \(\sH_V\), then \(\omega\) no longer has an active causal effect on \(A\). \definitionref{def:active_causal_effect_post-intervention} captures this property based on the causal kernels in the original causal space \((\Omega,\sH,\PP,\KK)\). If we let \(V=\emptyset\), then we recover the original definition of (active) causal effects in \definitionref{def:active_causal_effect}, since, by \axiomref{def:causal_space}(i), we have \(K_\emptyset(\omega,A)=\PP(A)\). 

\section{Quantifying measures of causal effects}\label{sec:quantifying_measures}
In this section, we move from existence of causal effects to their quantification by proposing \emph{mean effect} and \emph{maximum effect} scores. In \sectionref{subsec:causal_effect_events}, we consider causal effects on events, and in \sectionref{subsec:causal_effect_sigma_algebras}, we quantify causal effects on \(\sigma\)-algebras via a difference measure between distributions. We focus only on quantifying the vanilla marginal effects (corresponding to \definitionref{def:active_causal_effect}), but it is clear that the same ideas and tools can be extended to quantifying conditional and post-intervention causal effects. 

\subsection{Quantifying causal effects on events}\label{subsec:causal_effect_events}
We first quantify causal effects on events.
We compare the observational probability \(\PP\) with the interventional measure \(\iPP{U}{\QQ}\) through some scale function \(f\), whose role is to specify the level of emphasis on changes near the boundaries (rare or near-certain events).
\begin{definition}[Mean effect score]\label{def:mean_effect_score}Given an intervention set \(U\subseteq T\), an interventional measure \(\QQ\) on \((\Omega,\sH_U)\),
and an event \(A\in\sH\), the \emph{mean effect score} of an intervention on \(U\) via \(\QQ\) associated with a
scale function \(f:[0,1]\to[-\frac{1}{2},\frac{1}{2}]\) (see below) is defined as
\[\iS{U}{\QQ}{f}(A)=f(\iPP{U}{\QQ}(A))-f(\PP(A)).\]
\end{definition}
This score yields a signed measure of the causal effect of intervening on \(U\) for event \(A\) with the measure \(\QQ\). The sign indicates whether the intervention raises or lowers \(P(A)\), and \(f\) determines how changes are weighted across different baseline levels.

The scale function \(f\) should be chosen to reflect the needs of applications, in particular how strongly one wishes to weight probability changes near 0 or 1 compared with changes around \(\frac12\). We require \(f\) to be non-decreasing and symmetric around \(\frac{1}{2}\), with \(f(0)=-\tfrac12\), \(f(\tfrac12)=0\), and \(f(1)=\tfrac12\). These constraints satisfy three desiderata: (i) the score reflects both direction and magnitude of the shift between \(\PP(A)\) and \(K_U(\omega,A)\); (ii) it is normalised to \([0,1]\); and (iii) it treats \(A\) and \(A^c\) symmetrically over \([0,1]\). If one regards a given absolute probability change as equally consequential regardless of the original \(\PP(A)\), a natural choice is the linear map \[f_1(x)=x-\tfrac12,\] in which case \(\iS{U}{\QQ}{f}(A)=\iPP{U}{\QQ}(A)-\PP(A)\) (raw difference). If increases from virtually impossible to merely rare (e.g., \(0\!\to\!0.05\)) should count more than equal-sized increases around \(1/2\) (e.g., \(0.50\!\to\!0.55\)), we prefer \(f\) with a small slope around \(1/2\) and large slopes near the boundaries. One possible such choice is
\[
f_2(x)\;=\;\frac{\sinh\!\big(x-\tfrac12\big)}{2\,\sinh(\tfrac12)},
\]
which satisfies the boundary and symmetry conditions and has derivative \[f_2'(x)=\frac{\cosh(x-1/2)}{2\sinh(1/2)},\] minimal at \(x=\tfrac12\) and increasing as \(x\) approaches \(0\) or \(1\). 

Let us go back to our running example to illustrate the mean effect score. 
\begin{example}\label{ex:mean_effect_score_event}
Recall again that \(A=\{\omega_\pay=1000\}\). Let \(\QQ^Y,\QQ^N\) be measures such that \(\QQ^Y_{\ins}=\delta_Y\) and \(\QQ^N_{\ins}=\delta_N\). Using \tableref{tab:insurance_observational,tab:insurance_yes,tab:insurance_no}, the mean
effect scores for \(f_1\) and \(f_2\) with \(\QQ^Y\)
are respectively
\[
\iS{\ins}{\QQ^Y}{f_1}(A)=-0.00625, \qquad
\iS{\ins}{\QQ^Y}{f_2}(A)\approx-0.00675.
\]
The negative values of \(\iS{\ins}{\QQ^Y}{f_1}(A)\) and \(\iS{\ins}{\QQ^Y}{f_2}(A)\) indicate that buying insurance has the effect of reducing the probability of \(A\), with \(f_2\) putting more emphasis on the change than the raw difference. Of course, the difference is mild, but the scale function can be chosen to emphasise the effects on the boundaries more dramatically. 
With \(\QQ^N\), the scores are
\[
\iS{\ins}{\QQ^N}{f_1}(A)=0.00875, \qquad
\iS{\ins}{\QQ^N}{f_2}(A)\approx0.00942.
\]
These scores indicate not only that the causal effect of not buying insurance is an increase in the probability of \(A\), but also that the magnitude of the effects is greater than that of \(\QQ^Y\). 

\end{example}
While the mean effect score averages the impact of the intervention with respec to \(\QQ\), 
one may also be interested in the largest deviation induced by the intervention within a given event. This leads to the following definition of \emph{maximum effect score}.
\begin{definition}[Maximum effect score]\label{def:maximum_effect_score}
    Given an intervention set \(U\subseteq T\),
and events \(A\in\sH\), \(B\in\sH_U\), let us denote the outcome in \(B\) which leads to the maximal change in the probability of \(A\) under intervention on \(\sH_U\) as
\[\omega_{\max}=\argmax_{\omega\in B}\left\lvert f(K_U(\omega,A))-f(\PP(A))\right\rvert.\]
Then, the \emph{maximum effect score} associated with a
scale function \(f\) is defined by

\[\cS^{U,\max_B}_f(A)=f(K_U(\omega_{\max},A))-f(\PP(A)).\]
\end{definition}

\subsection{Quantifying causal effects on \texorpdfstring{\(\sigma\)}{sigma}-algebras}\label{subsec:causal_effect_sigma_algebras}
In this section, we consider causal effects on  \(\sigma\)-algebras \(\sF\subseteq\sH\), instead of individual events. In this case, we are not comparing two numbers, but a pair of distributions, which means that we need a way of comparing distributions. 

Let \(D_\sF\) be a measure of difference between two probability measures on \(\sF\). We do not, in general, require \(D_\sF\) to be a metric, but we require \(D_\sF(\cdot,\cdot)\) to take values in a normed vector space \((F,\lVert\cdot\rVert_F)\). In particular, it can take negative values and need not be a single value. In general, it can be multi-dimensional or even infinite-dimensional, and should be chosen according to the needs in practice, such as weighted sum of different aspects of the distributions, or the kernel mean embedding \citep{muandet2017kernel}.

We first define the mean effect score on \(\sigma\)-algebras. 
\begin{definition}[Mean effect score II]\label{def:mean_effect_score_sigma}Given a sub-\(\sigma\)-algebra \(\sF\) of \(\sH\), an intervention set \(U\subseteq T\), and an interventional measure \(\QQ\) on \((\Omega,\sH_U)\), the mean effect score associated with \(D_\sF\) is defined as
\[\iS{U}{\QQ}{D_\sF}=D_\sF(\iPP{U}{\QQ},\PP).\]
\end{definition}
In words, \(\iS{U}{\QQ}{D_\sF}\) compares the \(D_\sF\)-difference between the intervention measure \(\iPP{U}{\QQ}\) and the observational measure \(\PP\). In the following example, we illustrate the mean effect score with a particular choice of \(D_\sF\), namely, the difference in means. We show that, by considering the causal effect of a treatment in the causal space obtained by the control, and by choosing the difference measure \(D_\sF\) as the difference in means, we recover the ATE. 
\begin{example}[ATE]
For measures \(\mu\) and \(\nu\) on \(\sF\), define \(D_\sF\) as the difference in means:
\[D_\sF(\mu,\nu)=\int\omega\mu(d\omega)-\int\omega\nu(d\omega).\]

Let \(W\in\{0,1\}\) be a binary treatment variable, and \(Y\) the outcome variable. Let \(\sF=\sigma(Y)\). Consider an intervened causal space \((\Omega,\sH,\iPP{W}{\delta_0},\iKK{W}{\delta_0})\), obtained by intervening with \(W=0\). In this new causal space, the measure \(\iPP{W}{\delta_0}\) is
the law of \(Y\) under no treatment.
Now let us consider intervening further on \((\Omega,\sH,\iPP{W}{\delta_0},\iKK{W}{\delta_0})\), with the measure \(\QQ=\delta_1\). Then, by \definitionref{def:interventions}, we have
\[\iK{W}{\delta_0}{W}(1,\cdot)=K_W(1,\cdot),\]
which implies
\[(\iPP{W}{\delta_0})^{\textnormal{do}(W,\delta_1)}=\iPP{W}{\delta_1},\]
i.e., sequential interventions on the same treatment variable \(W\) with \(W=0\) first then \(W=1\) is the same as intervening with \(W=1\) in the original causal space. 

Hence, the difference-in-means mean effect
score corresponding to the intervention on \(\sH_W\) with \(\QQ=\delta_1\) in the new causal space \((\Omega,\sH,\iPP{W}{\delta_0},\iKK{W}{\delta_0})\) is
\[
\iS{U}{\QQ}{D_\sF}
= \EE_{Y\sim(\iPP{W}{\delta_1})^{\textnormal{do}(W,\delta_1)}}[Y]-\EE_{Y\sim\iPP{W}{\delta_0}}[Y]
=\EE_{Y\sim\iPP{W}{\delta_1}}[Y]-\EE_{Y\sim\iPP{W}{\delta_0}}[Y],
\]
precisely recovering the ATE with our definitions. By considering conditional causal effects with the same \(D_\sF\), it is straightforward to derive the CATE as a special case as well. 
\end{example}
Finally, let us give an analogue of \definitionref{def:maximum_effect_score}, i.e., the maximum effect score on \(\sigma\)-algebras. Recall that the difference measure \(D_\sF\) takes values in a normed space \((F,\lVert\cdot\rVert_F)\). 
\begin{definition}[Maximum effect score II]\label{def:maximum_effect_score_sigma}
Given a sub-\(\sigma\)-algebra \(\sF\) of \(\sH\), an intervention set \(U\subseteq T\) and an event \(B\in\sH_U\), let us denote the outcome in \(B\) leading to the maximal change in the distributional difference under intervention on \(\sH_U\) as
\[\omega_{\max}=\argmax_{\omega\in B}\lVert D_\sF(\iPP{U}{\delta_\omega},\PP)\rVert_F.\]
Then the maximum effect score associated with a measure \(D_\sF\) is defined as
\[\cS^{U,\max}_{D_\sF}=D_\sF(\iPP{U}{\delta_{\omega_{\max}}},\PP).\]

\end{definition}
The maximal distributional change is defined with respect to the norm in the normed space \((F,\lVert\cdot\rVert_F)\). If \(F=\RR\), then this simply boils down to the magnitude of the difference, and if \(F\) is a reproducing kernel Hilbert space, then it is the corresponding norm in this space. 

We return to our running example one last time to illustrate the distributional causal effects on the \(\sigma\)-algebra generated by the payment variable. 
\begin{example}
Intuitively, purchasing insurance increases the traveller's expected
payment, but at the same time dramatically reduces the variance of
this payment. Our definition can be used to verify this intuition.
Consider the causal effect on \(\sF=\sH_\pay\) of intervening on insurance with the delta measure \(\QQ_\ins=\delta_Y\), i.e., the causal effect of buying insurance. Let \(D_\sF^m\) and  \(D_\sF^v\) be the difference in the means and in the variances respectively. From the last column of \tableref{tab:insurance_observational,tab:insurance_yes}, the mean effect scores are respectively
\[\iS{U}{\QQ}{D_\sF^m}=8.45,\qquad \iS{U}{\QQ}{D_\sF^v}=-6244.5975.\]

This shows that, by buying insurance, the traveller loses \(8.45\) CHF on average, but completely removes the probability of having to pay \(1000\) CHF, reflected in the big reduction in the variance. This demonstrates that, when considering causal effects on distributions on \(\sigma\)-algebras \(\sF\), it is important to consider different aspects of the causal effect. 

\end{example}

\section{Conclusion}\label{sec:conclusion}
In this paper, we developed a measure-theoretic account of causal effects. Rather than starting from random variables, we define causal effects on events (and then on \(\sigma\)-algebras) within the causal space framework. This shift is motivated by (i) modern domains such as images and language models, where the raw variables (pixels, tokens) are often not the right semantic units for causal questions, and (ii) by drawing an analogy with the development of central notions such as independence in probability theory. 

Our first contribution is a family of binary (yes/no) notions of causal effect that can be stated at a fine granularity (individual outcomes \(\omega\) or events \(B\) affecting an event \(A\) or a \(\sigma\)-algebra \(\sF\)), together with conditional and post-intervention variants. We also established structural links to (conditional) independence under interventional measures. 

Our second contribution is to move from existence to quantification. For event-level targets we proposed mean and maximum effect scores that compare \(\PP(A)\) and \(\iPP{U}{\QQ}(A)\) through a user-chosen scale function, allowing the scoring to emphasize changes near the boundaries (rare or near-certain events) or to treat absolute probability changes uniformly. For \(\sigma\)-algebras, we proposed mean and maximum scores based on a generic \say{difference} functional \(D_\sF\) valued in a normed space, enabling multi-aspect and even infinite-dimensional summaries (e.g., via kernel mean embeddings). Importantly, standard treatment-effect objects---such as the ATE---are recovered as special cases of these definitions.

Finally, we stress that our focus here is on population-level definitions rather than inference: we did not develop identification results, estimation procedures or hypothesis tests for these effect notions from finite samples. Natural next steps include (i) identifying conditions under which the proposed quantities are estimable from observational or interventional data, (ii) designing tests for event-level and \(\sigma\)-algebra–level effects.

\acks{We are very grateful to Tobias Wegel, Florian Dorner, Julius von K\"ugelgen, Luigi Gresele, Fanny Yang and Thomas Icard for helpful discussions. JP is supported by SNSF Grant 218343.}

\bibliography{ref}

\clearpage
\appendix

\section{More details on related works}\label{sec:related}
In this section, we extend \sectionref{subsec:treatment_effect} to present the definitions of all the various definitions of treatment effects used in the potential outcomes framework. 

\subsection{Binary treatment}\label{subsec:binary_treatment}
We first introduce the potential outcomes framework in the case of a binary treatment, as in \sectionref{subsec:treatment_effect}. Let \((\Omega,\sH,\PP)\) be the underlying probability space, where we interpret \(\omega\in\Omega\) as \say{units}, e.g., patients in a study. Let \(W:\Omega\to\{0,1\}\) be a binary treatment variable, \(\bX\) a (vector of) covariate variable(s), and \(Y\), \(Y_0\) and \(Y_1\) the observed and potential outcomes under control and treatment respectively. The most elementary notion of causal effect is at the level of the \emph{individual} \(\omega\). The \emph{individual treatment effect} (ITE) is defined as
\[\text{ITE}(\omega)=Y_1(\omega)-Y_0(\omega).\]
This is clearly unidentifiable from observational data, since we do not observe both \(Y_1\) and \(Y_0\) for the same unit. 

As presented in \sectionref{subsec:treatment_effect}, the \emph{average treatment effect} (ATE) is defined as
\[\text{ATE}=\EE[Y_1-Y_0].\]
The ATE is identifiable from observations, and there are many ways of estimating this quantity using observational data. 

Next is the \emph{distributional treatment effect} (DTE), which is of interest if we are interested in the impact of the treatment beyond the expectation. Let \(P_{Y_0}\) and \(P_{Y_1}\) be the distributions of \(Y_0\) and \(Y_1\) respectively, and let \(D\) be some metric between distributions. Then the DTE associated with \(D\) is given by
\[\text{DTE}_D=D(P_{Y_0},P_{Y_1}).\]
Here, the choice of \(D\) depends on the aspect of the distributions that we aim to capture. For example, if \(D\) simply captures the difference in the means of the two distributions, then we recover the ATE. If it is designed to capture the difference in the quantiles, then we obtain the \emph{quantile treatment effect}. 

The conditional versions of the above two definitions are the \emph{conditional average treatment effect} (CATE) and the \emph{conditional distributional treatment effect} (CDTE):
\begin{alignat*}{2}
    \text{CATE}(\bx)&=\EE[Y_1-Y_0\mid\bX=\bx]\\
    \text{CDTE}_D(\bx)&=D(P_{Y_1\mid\bX=\bx},P_{Y_0\mid\bX=\bx}).
\end{alignat*}
The \emph{average treatment effect on the treated} (ATT) is used when we only care about the treatment effect on those who were actually treated. It is defined as
\[\text{ATT}=\EE[Y_1-Y_0\mid W=1].\]
The \emph{local average treatment effect} (LATE), or the average treatment effect on the compliers, is used when there is the issue of imperfect compliance, and is defined as
\[\text{LATE}=\EE[Y_1-Y_0\mid W_1>W_0].\]
The \emph{marginal average treatment effect} (MATE) is used when the resistance to treatment is quantified via a variable called \(V\), and is defined at a particular resistance level \(v\) as
\[\text{MATE}=\EE[Y_1-Y_0\mid V=v],\qquad v\in(0,1).\]
It is easy to extend the ATT, LATE and MATE to take into account treatment effect heterogeneity by conditioning on the covariates \(\bX\), or to distributional effects rather than average. 

\subsection{Continuous treatment}\label{subsec:continuous_treatment}
If the treatment variable \(W\) is multi-valued or even continuous instead of being binary, then instead of having a well-defined control level \(W=0\), one is free to choose any two treatment levels to compare. Let \(w_1\) and \(w_0\) be two possible values of \(W\). Then the ATE, also called the \emph{dose-response function} \citep{imbens2000role}, between \(w_1\) and \(w_0\) is defined as
\[\text{ATE}(w_0,w_1)=\EE[Y_{w_1}-Y_{w_0}],\]
with the analogous definitions for DTE, CATE, CDTE, etc. One can also consider the \emph{derivative effect} \citep{galvao2015uniformly} at a particular value \(w\) of treatment:
\[\tau(w)=\frac{d\EE[Y_w]}{dw}\]
to capture the effect of a small (infinitesimal) change in the treatment. One is also free to condition this on the covariates \(\bX\) to take treatment effect heterogeneity into account. 

\subsection{Total causal effect}\label{subsec:total_causal_effect}
The total causal effect in the SCM framework \citep[p.91, Definition 6.12]{peters2017elements} is slightly different, in that the baseline is not some control treatment, but the observational distribution. This is more aligned with our approach, but it is still only defined at the level of random variables. It is defined as follows. 
\begin{quote}
    There is a total causal effect from \(X\) to \(Y\) if and only if \(X\) and \(Y\) are not independent under the interventional distribution obtained after intervening on \(X\) to change it to a random variable \(\tilde{N}_X\). 
\end{quote}
\citet[p.70, Definition 3.2.1]{pearl2009causality} defines the \say{causal effect} as simply the interventional distribution, which is not the view that we adopt in this work. 

\subsection{Parallels with independence}\label{subsec:independence}
As mentioned in the introduction (\sectionref{sec:introduction}), the way we build up the theory of causal effects is analogous to how the theory of independence is developed in probability theory. First, a binary notion of independence is given, on the level of events. Given a probability space \((\Omega,\sH,\PP)\), we say that events \(A,B\in\sH\) are \emph{independent} if
\[\PP(A\cap B)=\PP(A)\PP(B).\]
Then, we say that \(\sigma\)-algebras \(\sF,\sG\subseteq\sH\) are independent if, for all \(A\in\sF\) and all \(B\in\sG\), \(A\) and \(B\) are independent. Finally, we say that two random variables are independent if the \(\sigma\)-algebras they generate are independent. 

Let us now consider conditioning. Let \(G\in\sH\) be an event, and \(\sG\subseteq\sH\) a \(\sigma\)-algebra. Then we say that \(A\) and \(B\) are conditionally independent given \(G\) if \(\PP(G)>0\) and
\[\PP_G(A\cap B)=\PP_G(A)\PP_G(B).\]
Similarly, we say that \(A\) and \(B\) are conditionally independent given \(\sG\) if, for \(\PP\)-almost all \(\omega\in\Omega\),
\[\PP_\sG(\omega,A\cap B)=\PP_\sG(\omega,A)\PP_\sG(\omega,B).\]
While (conditional) independence has a crisp, universal definition, the nature/strength of (conditional) dependence has no single canonical scalar accepted across contexts. Instead, many measures exist for different purposes, for example the Pearson correlation coefficient \citep{pearson1896vii,yule1907theory} for linear dependence, Kendall's \(\tau\) \citep{kendall1938new,martin2005testing} or Spearman's \(\rho\) \citep{spearman1904proof,schmid2007multivariate} for rank dependence, and mutual information \citep{shannon1948mathematical,wyner1978definition}, f-divergences \citep{ali1966general,wang2024generalization} or Hilbert-Schmidt (conditional) independence criterion \citep{gretton2005measuring,park2020measure} for general, distributional dependence. 

\section{No \& dormant causal effect}\label{sec:no_dormant}
In the main body of the paper, we only treated active causal effects (\definitionref{def:active_causal_effect}), which compares the causal kernel \(K_U(\omega,A)\) with the observational measure \(\PP\). In this section, we extend the definitions given in \sectionref{sec:binary} to the stronger notion of \emph{no causal effect} and \emph{dormant causal effect}. 

We start with the vanilla version of no causal effect. 
\begin{definition}[No causal effect]\label{def:no_causal_effect}
    Let us take a causal space \((\Omega,\sH,\PP,\KK)\), an intervention set \(U\subseteq T\), an outcome \(\omega\in\Omega\), events \(A,B\in\sH\) and a sub-\(\sigma\)-algebra \(\sF\subseteq\sH\). 
    \begin{enumerate}[(i)]
        \item If \(K_S((\omega_{S\cap U},\omega'_{S\setminus U}),A)=K_{S\setminus U}(\omega'_{S\setminus U},A)\) for all \(S\in\sP(T)\) and all \(\omega'_{S\setminus U}\in\Omega_{S\setminus U}\), then we say that \(\omega\) has \emph{no \(U\)-causal effect on \(A\)}.

        \item If no \(\omega\in B\) has a \(U\)-causal effect on \(A\), then we say that \(B\) has no \(U\)-causal effect on \(A\). 
    \end{enumerate}
    We say that \(\omega\) (resp.\ \(B\)) has no \(U\)-causal effect on \(\sF\) if \(\omega\) (resp.\ \(B\)) has no \(U\)-causal effect on any \(A\in\sF\). 
\end{definition}
The notion of \emph{no causal effect} is stronger than \emph{no active causal effect}---indeed, it is easy to show that if \(\omega\) has no causal effect on \(A\), then it does not have an active causal effect, but not vice versa. The case in which \(\omega\) has a causal effect but not an active causal effect is precisely the definition of \emph{dormant causal effect}, defined in the following.
\begin{definition}[Dormant causal effect]\label{def:dormant_causal_effect}
    Let us take a causal space \((\Omega,\sH,\PP,\KK)\), an intervention set \(U\subseteq T\), an outcome \(\omega\in\Omega\), events \(A,B\in\sH\) and a sub-\(\sigma\)-algebra \(\sF\subseteq\sH\). 
    \begin{enumerate}[(i)]
        \item If \(\omega\) has a \(U\)-causal effect on \(A\) (in the sense of \definitionref{def:no_causal_effect}) and does not have an active \(U\)-causal effect on \(A\) (in the sense of \definitionref{def:active_causal_effect}), we say that \(\omega\) has a \emph{dormant \(U\)-causal effect on \(A\)}.
        \item If \(B\) does not have an active \(U\)-causal effect on \(A\) but there exists some \(\omega\in B\) that has a dormant \(U\)-causal effect on \(A\), we say that \(B\) has a dormant \(U\)-causal effect on \(A\). 
    \end{enumerate}
    We say that \(\omega\) (resp.\ \(B\)) has a dormant \(U\)-causal effect on \(\sF\) if \(\omega\) (resp.\ \(B\)) has no active \(U\)-causal effect on any \(A\in\sF\) but has a dormant \(U\)-causal effect on some \(A\in\sF\). 
\end{definition}
In other words, the causal effect is dormant if intervening on \(U\) alone leaves \(\PP(A)\) unchanged, yet there exists some other set \(S\) such that intervening on \(U\cup S\) differs from intervening on \(S\) alone. 

Next are the conditional versions. Recall that we had the conditional version of active causal effects in \sectionref{subsec:conditional_causal_effect}. We continue to use the notation \(\iPP{U}{\delta_\omega}(\cdot)\) for \(K_U(\omega,\cdot)\). We first condition on an event \(G\in\sH\), as we did in \definitionref{def:active_causal_effect_conditional_event}. 
\begin{definition}[Conditional no causal effect I]\label{def:no_causal_effect_conditional_event}
    Take a causal space \((\Omega,\sH,\PP,\KK)\), an intervention set \(U\subseteq T\), an outcome \(\omega\in\Omega\), events \(A,B,G\in\sH\) and a sub-\(\sigma\)-algebra \(\sF\subseteq\sH\). 
    \begin{enumerate}[(i)]
        \item For each \(\omega'_{S\setminus U}\in\Omega_{S\setminus U}\), write \(\omega'=(\omega'_{S\setminus U},\omega_{T\setminus(S\setminus U)})\in\Omega\) for the outcome obtained by replacing the \(S\setminus U\) components of \(\omega\) with \(\omega'_{S\setminus U}\). 
        
        If, for all \(S\subseteq T\) and all \(\omega'_{S\setminus U}\in\Omega_{S\setminus U}\), we have \(\iPP{S}{\delta_{\omega'}}(G)>0\), \(\iPP{S\setminus U}{\delta_{\omega'}}(G)>0\) and \(\iPP{S}{\delta_{\omega'}}_G(A)=\iPP{S\setminus U}{\delta_{\omega'}}_G(A)\), then we say that \(\omega\) has no \(U\)-causal effect on \(A\) conditioned on \(G\). 
        \item If no \(\omega\in B\) has a \(U\)-causal effect on \(A\) conditioned on \(G\), then we say that \(B\) has no \(U\)-causal effect on \(A\) conditioned on \(G\). 
    \end{enumerate}
    We say that \(\omega\) (resp.\ \(B\)) has no \(U\)-causal effect on \(\sF\) conditioned on \(G\) if \(\omega\) (resp.\ \(B\)) has no \(U\)-causal effect on any \(A\in\sF\) conditioned on \(G\). 
\end{definition}
In \definitionref{def:no_causal_effect_conditional_event}(i), we require \(G\) to have positive measure under \(\iPP{S}{\delta_{\omega'}}\) and \(\iPP{S\setminus U}{\delta_{\omega'}}\) for \emph{all} \(S\subseteq T\) and \emph{all} \(\omega'_{S\setminus U}\in\Omega_{S\setminus U}\). Otherwise, the existence of conditional causal effect cannot be determined. 

Next, we give the conditional version of dormant causal effect, given an event \(G\in\sH\). 
\begin{definition}[Conditional dormant effect I]\label{def:dormant_causal_effect_conditional_event}
    Let us take a causal space \((\Omega,\sH,\PP,\KK)\), an intervention set \(U\subseteq T\), an outcome \(\omega\in\Omega\), events \(A,B,G\in\sH\) and a sub-\(\sigma\)-algebra \(\sF\subseteq\sH\). 
    \begin{enumerate}[(i)]
        \item If \(\omega\) has a \(U\)-causal effect on \(A\) conditioned on \(G\) (in the sense of \definitionref{def:no_causal_effect_conditional_event}) and does not have an active \(U\)-causal effect on \(A\) conditioned on \(G\) (in the sense of \definitionref{def:active_causal_effect_conditional_event}), we say that \(\omega\) has a \emph{dormant \(U\)-causal effect} on \(A\) given \(G\).
        \item If \(B\) does not have an active \(U\)-causal effect on \(A\) conditioned on \(G\) but there exists some \(\omega\in B\) that has a dormant \(U\)-causal effect on \(A\) conditioned on \(G\), we say that \(B\) has a dormant \(U\)-causal effect on \(A\) conditioned on \(G\). 
    \end{enumerate}
    We say that \(\omega\) (resp.\ \(B\)) has a dormant \(U\)-causal effect on \(\sF\) conditioned on \(G\) if \(\omega\) (resp.\ \(B\)) has no active \(U\)-causal effect on any \(A\in\sF\) conditioned on \(G\) but has a dormant \(U\)-causal effect on some \(A\in\sF\) conditioned on \(G\). 
\end{definition}
We also give the corresponding definitions when we condition on a \(\sigma\)-algebra \(\sG\). 
\begin{definition}[Conditional no causal effect II]\label{def:no_causal_effect_conditional}
    Take a causal space \((\Omega,\sH,\PP,\KK)\), an intervention set \(U\subseteq T\), an outcome \(\omega\in\Omega\), events \(A,B\in\sH\) and sub-\(\sigma\)-algebras \(\sF,\sG\subseteq\sH\). 
    \begin{enumerate}[(i)]
        \item For each \(\omega'_{S\setminus U}\in\Omega_{S\setminus U}\), write \(\omega'=(\omega'_{S\setminus U},\omega_{T\setminus(S\setminus U)})\in\Omega\) for the outcome obtained by replacing the \(S\setminus U\) components of \(\omega\) with \(\omega'_{S\setminus U}\). 
        
        If, for all \(S\subseteq T\) and all \(\omega'_{S\setminus U}\in\Omega_{S\setminus U}\), we have that \(\iPP{S}{\delta_{\omega'}}\) and \(\iPP{S\setminus U}{\delta_{\omega'}}\) are absolutely continuous with respect to each other on \(\sG\), and \(\iPP{S}{\delta_{\omega'}}_\sG(\tilde{\omega},A)=\iPP{S\setminus U}{\delta_{\omega'}}_\sG(\tilde{\omega},A)\) for \(\iPP{S}{\delta_{\omega'}}\)-almost all \(\tilde{\omega}\in\Omega\), then we say that \(\omega\) has no \(U\)-causal effect on \(A\) conditioned on \(\sG\). 
        \item If no \(\omega\in B\) has a \(U\)-causal effect on \(A\) conditioned on \(\sG\), then we say that \(B\) has no \(U\)-causal effect on \(A\) conditioned on \(\sG\). 
    \end{enumerate}
    We say that \(\omega\) (resp.\ \(B\)) has no \(u\)-causal effect on \(\sF\) conditioned on \(\sG\) if \(\omega\) (resp.\ \(B\)) has no active \(U\)-causal effect on any \(A\in\sF\) conditioned on \(\sG\). 
\end{definition}
Just like in \definitionref{def:active_causal_effect_conditional}, we require the measures \(\iPP{S}{\delta_{\omega'}}\) and \(\iPP{S\setminus U}{\delta_{\omega'}}\) to be absolutely continuous on \(\sG\), but here, for \emph{all} \(S\subseteq T\) and \emph{all} \(\omega'_{S\setminus U}\in\Omega_{S\setminus U}\). Otherwise, the existence of conditional causal effect cannot be determined. 

Next, we give the conditional version of dormant causal effect, given a \(\sigma\)-algebra \(\sG\in\sH\). 
\begin{definition}[Conditional dormant effect II]\label{def:dormant_causal_effect_conditional}
    Take a causal space \((\Omega,\sH,\PP,\KK)\), an intervention set \(U\subseteq T\), an outcome \(\omega\in\Omega\), events \(A,B\in\sH\) and sub-\(\sigma\)-algebras \(\sF,\sG\subseteq\sH\). 
    \begin{enumerate}[(i)]
        \item If \(\omega\) has a \(U\)-causal effect on \(A\) conditioned on \(\sG\) (in the sense of \definitionref{def:no_causal_effect_conditional}) and does not have an active \(U\)-causal effect on \(A\) conditioned on \(\sG\) (in the sense of \definitionref{def:active_causal_effect_conditional}), we say that \(\omega\) has a \emph{dormant \(U\)-causal effect} on \(A\) given \(\sG\).
        \item If \(B\) does not have an active \(U\)-causal effect on \(A\) conditioned on \(\sG\) but there exists some \(\omega\in B\) that has a dormant \(U\)-causal effect on \(A\) conditioned on \(\sG\), we say that \(B\) has a dormant \(U\)-causal effect on \(A\) given \(\sG\). 
    \end{enumerate}
    We say that \(\omega\) (resp.\ \(B\)) has a dormant \(U\)-causal effect on \(\sF\) conditioned on \(\sG\) if \(\omega\) (resp.\ \(B\)) has no active \(U\)-causal effect on any \(A\in\sF\) conditioned on \(\sG\) but has a dormant \(U\)-causal effect on some \(A\in\sF\) conditioned on \(\sG\). 
\end{definition}
In the following remark, we collect some straightforward yet important properties of conditional causal effects. 
\begin{remark}
    \begin{enumerate}[(a)]
        \item If \(\omega\) has no \(U\)-causal effect on \(A\) conditioned on \(G\), then letting \(S=U\) in \definitionref{def:no_causal_effect_conditional}(i), we can see that
        \[\iPP{U}{\delta_\omega}_G(A)=\iPP{\emptyset}{\delta_\omega}_G(A)=\PP_G(A).\]
        According to \definitionref{def:active_causal_effect_conditional}, we have that \(\omega\) has no active \(U\)-causal effect on \(A\) conditioned on \(G\). Hence, \(\omega\) cannot have both no \(U\)-causal effect and active \(U\)-causal effect on \(A\) conditioned on \(G\). 
        
        Similarly, letting \(S=U\) in \definitionref{def:no_causal_effect_conditional}, we have a direct contradiction with the definition of active causal effects given in \definitionref{def:active_causal_effect_conditional}. Hence, \(\omega\) cannot have both no \(U\)-causal effect and active \(U\)-causal effect on \(A\) conditioned on \(\sG\). 
        \item For any \(U\subseteq T\), no \(\omega\in\Omega\) has a \(U\)-causal effect on any \(G\in\sH\) conditioned on \(G\). Indeed, for any \(S\subseteq T\) and any \(\omega'_{S\setminus U}\in\Omega_{S\setminus U}\), we have
        \[\iPP{S}{\delta_{\omega'}}_G(G)=1=\iPP{S\setminus U}{\delta_{\omega'}}_G(G).\]
        For any \(U\in\sP(T)\), no \(\omega\in\Omega\) has a \(U\)-causal effect on any event \(A\in\sG\) conditioned on \(\sG\). Indeed, for any event \(A\in\sG\), any \(S\subseteq T\) and any \(\omega'_{S\setminus U}\in\Omega_{S\setminus U}\), we have
        \[\iPP{S}{\delta_{\omega'}}_\sG(\tilde{\omega},A)=\ind_A(\tilde{\omega})=\iPP{S\setminus U}{\delta_{\omega'}}_\sG(\tilde{\omega},A)\]
        for \(\iPP{S}{\delta_{\omega'}}\)-all \(\tilde{\omega}\in\Omega\).
        \item Let \(U\in\sP(T)\) and \(\sF_1,\sF_2\) be sub-\(\sigma\)-algebras of \(\sH\). If \(\sF_1\subseteq\sF_2\) and \(\omega\) has no \(U\)-causal effect on \(\sF_2\) conditioned on \(G\) (or \(\sG\)), then it is clear that \(\omega\) has no \(U\)-causal effect on \(\sF_1\) conditioned on \(G\) (or \(\sG\)).
    \end{enumerate}
\end{remark}
Finally, we give the post-intervention versions of no and dormant causal effects. Recall that the active post-intervention causal effect was defined in \definitionref{def:active_causal_effect_post-intervention}. 
\begin{definition}[No post-intervention causal effect]\label{def:no_causal_effect_post-intervention}
    Take a causal space \((\Omega,\sH,\PP,\KK)\), subsets \(U,V\in\sP(T)\), an outcome \(\omega\in\Omega\), events \(A,B\in\sH\) and a \(\sigma\)-algebra \(\sF\subseteq\sH\). 
	\begin{enumerate}[(i)]
		\item For each \(S\subseteq T\) and \(\omega'_{(S\cup V)\setminus(U\setminus V)}\in\Omega\), write \(\omega'=(\omega'_{(S\cup V)\setminus(U\setminus V)},\omega_{T\setminus(S\cup V)\setminus(U\setminus V)})\in\Omega\) for the outcome obtained by replacing the \((S\cup V)\setminus(U\setminus V)\) components of \(\omega\) with \(\omega'_{(S\cup V)\setminus(U\setminus V)}\). 
        
        If
        \[K_{S\cup V}(\omega',A)=K_{(S\cup V)\setminus(U\setminus V)}(\omega',A)\]
        for all \(S\subseteq T\) and all \(\omega'_{(S\cup V)\setminus(U\setminus V)}\in\Omega_{(S\cup V)\setminus(U\setminus V)}\), then we say that \(\omega\) has no \(U\)-causal effect on \(A\) after intervening on \(\sH_V\). 
		\item If no \(\omega\in B\) has a \(U\)-causal effect on \(A\) after intervening on \(\sH_V\), then we say that \(B\) has no \(U\)-causal effect on \(A\) after intervening on \(\sH_V\). 
	\end{enumerate}
    We say that \(\omega\) (resp.\ \(B\)) has no \(U\)-causal effect on \(\sF\) after intervening on \(\sH_V\) if \(\omega\) (resp.\ \(B\)) has no \(U\)-causal effect on any \(A\in\sF\) after intervening on \(\sH_V\). 
\end{definition}
\begin{definition}[Dormant post-intervention causal effect]\label{def:dormant_causal_effect_post-intervention}
    Let us take a causal space \((\Omega,\sH,\PP,\KK)\), subsets \(U,V\subseteq T\), an outcome \(\omega\in\Omega\), events \(A,B\in\sH\) and a sub-\(\sigma\)-algebra \(\sF\subseteq\sH\). 
    \begin{enumerate}[(i)]
        \item If \(\omega\) has a \(U\)-causal effect on \(A\) after intervening on \(\sH_V\) (in the sense of \definitionref{def:no_causal_effect_post-intervention}) and does not have an active \(U\)-causal effect on \(A\) after intervening on \(\sH_V\) (in the sense of \definitionref{def:active_causal_effect_post-intervention}), we say that \(\omega_U\) has a \emph{dormant \(U\)-causal effect} on \(A\) after intervening on \(\sH_V\).
        \item If \(B\) does not have an active \(U\)-causal effect on \(A\) after intervening on \(\sH_V\) but there exists some \(\omega\in B\) that has a dormant \(U\)-causal effect on \(A\) after intervening on \(\sH_V\), we say that \(B\) has a dormant \(U\)-causal effect on \(A\) after intervening on \(\sH_V\). 
    \end{enumerate}
    We say that \(\omega\) (resp.\ \(B\)) has a \(U\)-causal effect on \(\sF\) after intervening on \(\sH_V\) if \(\omega\) (resp.\ \(B\)) has no active \(U\)-causal effect on any \(A\in\sF\) after intervening on \(\sH_V\) but has a dormant \(U\)-causal effect on some \(A\in\sF\) after intervening on \(\sH_V\). 
\end{definition}

\section{Marginalisation}\label{sec:marginalisation}
The notion of marginal causal spaces has not yet appeared in previous works (for marginalisations of SCMs, see \citet{evans2016graphs,evans2018margins,bongers2021foundations}), but it is a simple extension of marginal probability spaces. We formally define it here. 

Let \(T\) and \(\tilde{T}\) be index sets such that \(T\subseteq\tilde{T}\). Suppose that we have two causal spaces \(\cC\) and \(\tilde{\cC}\), with index sets \(T\) and \(\tilde{T}\) respectively:
\begin{equation*}
    \cC=(\times_{t\in T}\Omega_t,\otimes_{t\in T}\sE_t,\PP,\KK),\qquad\tilde{\cC}=(\times_{t\in\tilde{T}}\Omega_t,\otimes_{t\in\tilde{T}}\sE_t,\tilde{\PP},\tilde{\KK}),
\end{equation*}
where \(\KK=\{K_S:S\in\sP(T)\}\) and \(\tilde{\KK}=\{\tilde{K}_S:S\in\sP(\tilde{T})\}\) are the causal mechanisms. The measurable spaces \((\Omega_t,\sE_t)\) for \(t\in T\) are required to be the same for both \(\cC\) and \(\tilde{\cC}\). Write \(\Omega=\times_{t\in T}\Omega_t\), \(\tilde{\Omega}=\times_{t\in\tilde{T}}\Omega_t\) and \(\Omega^\setminus=\times_{t\in\tilde{T}\setminus T}\Omega_t\), so that \(\Omega\times\Omega^\setminus=\tilde{\Omega}\). Similarly, write \(\sH=\otimes_{t\in T}\sE_t\) and \(\tilde{\sH}=\otimes_{t\in\tilde{T}}\sE_t\). 

Recall that the probability measure \(\PP\) is a marginalisation of \(\tilde{\PP}\) if, for all \(A\in\sH\), we have \(\PP(A)=\tilde{\PP}(A\times\Omega^\setminus)\). We say that the causal mechanism \(\KK\) is a marginalisation of \(\tilde{\KK}\) if, for any \(A\in\sH\) and any \(S\in\sP(T)\), we have \(K_S(\omega_S,A)=\tilde{K}_S(\omega_S,A\times\Omega^\setminus)\) for all \(\omega_S\in\Omega_S\), where we recall that \(\Omega_S=\times_{t\in S}\Omega_t\). In other words, each of the causal kernels in the smaller causal space \(\cC\) is a marginalisation of the corresponding causal kernel in the larger causal space \(\tilde{\cC}\), for each \(\omega_S\in\Omega_S\). We say that \(\cC\) is a marginalisation of \(\tilde{\cC}\) if \(\PP\) is a marginalisation of \(\tilde{\PP}\) and \(\KK\) is a marginalisation of \(\tilde{\KK}\). 

Now, let us return to causal effects in \sectionref{sec:binary} and \appendixref{sec:no_dormant}. 
\begin{remark}\label{rem:marginalisation}
    \begin{enumerate}[(i)]
        \item The definitions of no causal effect (\definitionref{def:no_causal_effect,def:no_causal_effect_conditional_event,def:no_causal_effect_conditional,def:no_causal_effect_post-intervention}) are \emph{not} preserved by marginalisation, since they depend on all possible subsets \(S\) that one can intervene on. Specifically, suppose that there is some \(\tilde{S}\in\sP(\tilde{T})\) in the larger causal space \(\tilde{\cC}\) such that \(\tilde{K}_{\tilde{S}}(\omega,A)\neq\tilde{K}_{\tilde{S}\setminus U}(\omega,A)\) for some \(\omega\in\Omega\), but the marginalisation is such that \(\tilde{S}\subsetneq\sP(T)\). Suppose further that we do have \(K_S(\omega,A)=K_{S\setminus U}(\omega,A)\) for all \(S\in\sP(T)\) and all \(\omega\in\Omega\). Then \(\omega_U\) has no causal effect on \(A\) in the smaller causal space \(\cC\) but does have a causal effect on \(A\) in the larger causal space \(\tilde{\cC}\). 
        \item However, it can be trivially seen that
        \begin{enumerate}
            \item if \(\omega_U\) has a causal effect on \(A\) in the smaller causal space \(\cC\), then it will always have a causal effect on \(A\) in the larger causal space \(\tilde{\cC}\);
            \item if \(\omega_U\) has no causal effect on \(A\) in the larger causal space \(\tilde{\cC}\), then it will always have no causal effect on \(A\) in the smaller causal space \(\cC\). 
        \end{enumerate}
    \item Active causal effect (\sectionref{sec:binary}) is preserved by marginalisation. Indeed, if, for an event \(A\in\sH\), we have \(\tilde{K}_U(\omega,A\times\Omega^{\setminus})\neq\tilde{\PP}(A\times\Omega^{\setminus})\) for some \(\omega\in\Omega\), then by the definition of marginalisation, \(K_U(\omega,A)\neq\PP(A)\), and vice versa. 
    \end{enumerate}
\end{remark}

\section{Language model example}\label{sec:lm}
We follow the measure-theoretic, stochastic process view of language models \citep{cotterell2023formal,meister2023locally,du2023measure,du2024language}. Let \(\cV\) be a (finite) vocabulary of tokens, equipped with the power set \(\sigma\)-algebra \(2^\cV\). This vocabulary includes the \emph{end-of-sequence} token \texttt{EOS}. We consider two copies of the product measurable space \((\Omega^P,\sH^P)=\otimes_{n\in\NN}(\cV,2^\cV)\) and \((\Omega^O,\sH^O)=\otimes_{n\in\NN}(\cV,2^\cV)\) to which prompt and output text respectively belong. We suppose that the weights of the language model are frozen, and henceforth do not write the dependence on the weights explicitly. 

We consider the causal effects of the prompt on the probability over the output text. In language models, causal queries are often about properties of the generated text, such as topic or style, rather than about individual tokens. A variable-level definition can be unnatural in this setting. The raw variables are tokens, while the causal questions are often about higher-level properties of the whole output. Our event-level definitions allows such questions to be formulated directly as events in the output space, without requiring the introduction of pre-defined variables. 

We have the causal kernel \(K_P\), which is a transition probability kernel from \((\Omega^P,\sH^P)\) into \((\Omega^O,\sH^O)\), i.e., for each prompt \(\bp\in\Omega^P\), we have a probability measure \(K_P(\bp,\cdot)\) on the output text space \((\Omega^O,\sH^O)\) to obtain a language process \citep[Definition 2.2]{meister2023locally}. Further, we assume that the weights of the model are such that, for any prompt \(\bp\), the process is \emph{tight} \citep[Definition 3.2]{du2024language}, i.e., the stopping time \(\tau_\texttt{EOS}\), defined as the first time the end-of-sequence token \texttt{EOS} is reached, is almost surely finite: \(K_P(\bp,\{\tau_\texttt{EOS}<\infty\})=1\). 

Then the observational measure \(\PP\) on \((\Omega^O,\sH^O)\) is given via a probability measure \(\PP^P\) on the prompt space \((\Omega^P,\sH^P)\): for any \(A\in\sH^O\), we have
\[\PP(A)=\int\PP^P(d\bp)K_P(\bp,A).\]
Now, along with the notions introduced in the main body, we have all the ingredients for the study of causal effects of the prompt on the output text in language models. 

Consider events \(A,A_C,A_P\in\sH^O\), defined as
\begin{alignat*}{2}
    A&=\{\text{the generated text is about politics}\}\\
    A_C&=\{\text{the generated text is written in a politically conservative tone}\}\\
    A_P&=\{\text{the generated text is written in a politically progressive tone}\},
\end{alignat*}
with the relationships \(A_C,A_P\subseteq A\), \(A_C\cap A_P=\emptyset\) and \(A_C\cup A_P\subsetneq A\). With respect to the observational marginal measure \(\PP\), all of these events have negligible probabilities, i.e. \(\PP(A)\ll1\), since, among all possible texts, the observed proportion of political texts is very small. 

Suppose that \(B_P\in\sH^P\) is an event in the prompt space consisting of prompts that ask the model to write in a politically progressive tone. Then for each \(\bp\in B_P\), we will have \(K_P(\bp,A_P)\gg\PP(A_P)\), i.e., the probability that the generated text is written in a politically progressive tone will be much higher than the observational probability for this event, so \(B_P\) will have a large causal effect on the event \(A_P\). However, it will have negligible, if any, causal effect on the event \(A_C\), since the probability of \(A_C\) was negligible anyway. 

Now, suppose that \(B\in\sH^P\) is an event that consists of all prompts that ask for a political output; in particular, we have \(B_P\subsetneq B\). Then the conditional probability \(\PP_B(A)\) will be close to 1, and \(\PP_B(A_P)\) and \(\PP_B(A_C)\) will also be no longer negligible. Now, if we intervened with a prompt \(\bp\) in \(B_P\), then \(\iPP{P}{\delta_{\bp}}_B=\iPP{P}{\delta_{\bp}}\), since \(\bp\in B_P\subset B\), and \(\PP_B(A_C)\gg\iPP{P}{\delta_{\bp}}_B(A_C)\), since asking the model to produce a politically progressive text will significantly reduce the probability of the event \(A_C\) that the output text is written in a conservative tone. 

This example illustrates why an event-level notion of causal effect is useful. In language model settings, the causal question of interest are typically higher-level properties of the generated text, rather than the individual tokens. By formulating causal effects directly on events in the output space, our framework makes it possible to study such semantic effects.

\section{Proofs}\label{sec:proofs}
\printlemma{lem:independence}
\begin{proof}
    Since \(\Omega\) has no active causal effect on \(A\), we have \(K_U(\omega,A)=\PP(A)\) for all \(\omega\in\Omega\). Then see that, for any \(B\in\sH_U\), the intervention measure \(\iPP{U}{\QQ}\) on the intersection \(A\cap B\) is
    \begin{alignat*}{3}
        \iPP{U}{\QQ}(A\cap B)&=\int\QQ(d\omega)K_U(\omega,A\cap B)&&\text{by \definitionref{def:interventions}}\\
        &=\int\QQ(d\omega)\ind_B(\omega)K_U(\omega,A)\quad&&\text{by \axiomref{def:causal_space}(ii), since }B\in\sH_U\\
        &=\int\QQ(d\omega)\ind_B(\omega)\PP(A)&&\text{since }K_U(\omega,A)=\PP(A)\text{ for all }\omega\in B\\
        &=\QQ(B)\PP(A).
    \end{alignat*}
    On the individual events \(A\) and \(B\), they are
    \[\iPP{U}{\QQ}(A)=\int\QQ(d\omega)K_U(\omega,A)=\int\QQ(d\omega)\PP(A)=\PP(A)\]
    and by \axiomref{def:causal_space}(ii),
    \[\iPP{U}{\QQ}(B)=\int\QQ(d\omega)K_U(\omega,B)=\int\QQ(d\omega)\ind_B(\omega)=\QQ(B).\]
    Hence, we have
    \[\iPP{U}{\QQ}(A\cap B)=\iPP{U}{\QQ}(A)\iPP{U}{\QQ}(B),\]
    i.e., \(A\) and \(B\) are independent under the measure \(\iPP{U}{\QQ}\). Since \(B\in\sH_U\) was arbitrary, we conclude that \(\sH_U\) and \(A\) are independent under \(\iPP{U}{\QQ}\). 
\end{proof}

\printprop{prop:conditional_independence}
\begin{proof}
    We present the proof in two parts, the first part (i) in the case of conditioning on an event \(G\), and the second part (ii) in the more involved case of conditioning on a \(\sigma\)-algebra \(\sG\). 
    \begin{enumerate}[(i)]
        \item Recall that \(G\) has positive measure under both \(\PP\) and \(\iPP{U}{\delta_\omega}\), so that the conditional probabilities \(\PP_G\) and \(\iPP{U}{\delta_\omega}_G\) are well-defined. Since \(\Omega\) has no active causal effect on \(A\) conditioned on \(G\), we have, for all \(\omega\in\Omega\), 
        \[\PP_G(A)=\iPP{U}{\delta_\omega}_G(A)=\frac{K_U(\omega,A\cap G)}{K_U(\omega,G)}.\tag{*}\]
        Hence see that, for any \(B\in\sH_U\), 
        \begin{alignat*}{3}
            \iPP{U}{\QQ}_G(A\cap B)&=\frac{1}{\iPP{U}{\QQ}(G)}\int\QQ(d\omega)K_U(\omega,A\cap B\cap G)&&\text{by \definitionref{def:interventions}}\\
            &=\frac{1}{\iPP{U}{\QQ}(G)}\int\QQ(d\omega)\ind_B(\omega)K_U(\omega,A\cap G)&&\text{by \axiomref{def:causal_space}(ii)}\\
            &=\frac{1}{\iPP{U}{\QQ}(G)}\int\QQ(d\omega)\ind_B(\omega)\PP_G(A)K_U(\omega,G)\quad&&\text{by (*)}\\
            &=\frac{\PP_G(A)}{\iPP{U}{\QQ}(G)}\int\QQ(d\omega)\ind_B(\omega)K_U(\omega,G)\\
            &=\frac{\PP_G(A)}{\iPP{U}{\QQ}(G)}\int\QQ(d\omega)K_U(\omega,G\cap B)&&\text{by \axiomref{def:causal_space}(ii)}\\
            &=\PP_G(A)\frac{\iPP{U}{Q}(G\cap B)}{\iPP{U}{\QQ}(G)}&&\text{by \definitionref{def:interventions}}\\
            &=\PP_G(A)\iPP{U}{\QQ}_G(B).
        \end{alignat*}
        But here, see that
        \begin{alignat*}{3}
            \iPP{U}{\QQ}_G(A)&=\frac{1}{\iPP{U}{\QQ}(G)}\int\QQ(d\omega)K_U(\omega,A\cap G)\\
            &=\frac{1}{\iPP{U}{\QQ}(G)}\int\QQ(d\omega)\PP_G(A)K_U(\omega,G)\qquad&&\text{by (*)}\\
            &=\frac{\PP_G(A)}{\iPP{U}{\QQ}(G)}\iPP{U}{\QQ}(G)\\
            &=\PP_G(A).
        \end{alignat*}
        Hence, combining with above, we have
        \[\iPP{U}{\QQ}_G(A\cap B)=\iPP{U}{\QQ}_G(A)\iPP{U}{\QQ}_G(B),\]
        meaning \(A\) and \(B\) are conditionally independent given \(G\) under the intervention measure \(\iPP{U}{\QQ}\). Since \(B\in\sH_U\) was arbitrary, \(A\) and \(\sH_U\) are conditionally independent given \(G\) under the measure \(\iPP{U}{\QQ}\).  
        \item Let us first see that, since \(\Omega\) has no causal effect on \(A\) conditioned on \(\sG\), we have, for all \(H\in\sG\) and all \(\omega\in\Omega\),
        \[\iEE{U}{\delta_\omega}\left[\ind_H\iPP{U}{\delta_\omega}_\sG(\cdot,A)\right]=\iEE{U}{\delta_\omega}\left[\ind_H\PP_\sG(\cdot,A)\right].\tag{**}\]
        Then for any \(H\in\sG\), 
        \begin{alignat*}{3}
            \iPP{U}{\QQ}(A\cap H)&=\int\QQ(d\omega)K_U(\omega,A\cap H)&&\text{\definitionref{def:interventions}}\\
            &=\int\QQ(d\omega)\iPP{U}{\delta_\omega}(A\cap H)\\
            &=\int\QQ(d\omega)\iEE{U}{\delta_\omega}\left[\ind_H\iPP{U}{\delta_\omega}_\sG(\cdot,A)\right]&&\text{tower property}\\
            &=\int\QQ(d\omega)\iEE{U}{\delta_\omega}\left[\ind_H\PP_\sG(\cdot,A)\right]&&\text{by (**)}\\
            &=\int\QQ(d\omega)\int\iPP{U}{\delta_\omega}(d\omega')\ind_H(\omega')\PP_\sG(\omega',A)\\
            &=\int\QQ(d\omega)\int K_U(\omega_U,d\omega')\ind_H(\omega')\PP_\sG(\omega',A)\\
            &=\iEE{U}{\QQ}\left[\ind_H\PP_\sG(\cdot,A)\right].
        \end{alignat*}
        Hence, for \(\iPP{U}{\QQ}\)-almost all \(\omega'\in\Omega\), we have \(\iPP{U}{\QQ}_\sG(\omega',A)=\PP_\sG(\omega',A)\). 

        Now, by \citet[p.10, Theorem I.2.17]{cinlar2011probability}, a positive function is measurable if and only if it is the limit of an increasing sequence of positive simple functions. Since the map \(\omega'\mapsto\PP_\sG(\omega',A):\Omega\to[0,1]\) is a \(\sG\)-measurable function, we can write it as a limit of positive \(\sG\)-simple functions, say \((f_n)_{n\in\NN}\) with
        \[f_n=\sum^{m_n}_{i_n=1}b_{i_n}\ind_{H_{i_n}}\]
        where \(b_{i_n}\in\RR\) and \(H_{i_n}\in\sG\). Then for any \(H\in\sG\) and any \(B\in\sH_U\),
        \begin{alignat*}{3}
            &\iEE{U}{\QQ}\left[\ind_H\iPP{U}{\QQ}_\sG(\cdot,A\cap B)\right]\\
            &=\iPP{U}{\QQ}(A\cap B\cap H)&&\text{tower property}\\
            &=\int\QQ(d\omega)K_U(\omega,A\cap B\cap H)&&\text{\definitionref{def:interventions}}\\
            &=\int\QQ(d\omega)\ind_B(\omega)K_U(\omega,A\cap H)&&\text{\axiomref{def:causal_space}(ii)}\\
            &=\int\QQ(d\omega)\ind_B(\omega)\iPP{U}{\delta_\omega}(A\cap H)\\
            &=\int\QQ(d\omega)\ind_B(\omega)\iEE{U}{\delta_\omega}\left[\ind_H\iPP{U}{\delta_\omega}_\sG(\cdot,A)\right]&&\text{tower property}\\
            &=\int\QQ(d\omega)\ind_B(\omega)\iEE{U}{\delta_\omega}\left[\ind_H\PP_\sG(\cdot,A)\right]&&\text{by (**)}\\
            &=\int\QQ(d\omega)\ind_B(\omega)\int\iPP{U}{\delta_\omega}(d\omega')\ind_H(\omega')\left(\lim_{n\to\infty}f_n(\omega')\right)\\
            &=\lim_{n\to\infty}\int\QQ(d\omega)\ind_B(\omega)\int K_U(\omega,d\omega')\ind_H(\omega')f_n(\omega')\enspace&&\text{monotone convergence}\\
            &=\lim_{n\to\infty}\sum^{m_n}_{i_n=1}b_{i_n}\int\QQ(d\omega)\ind_B(\omega)K_U(\omega,H\cap H_{i_n})\\
            &=\lim_{n\to\infty}\sum^{m_n}_{i_n=1}b_{i_n}\int\QQ(d\omega)K_U(\omega,H\cap H_{i_n}\cap B)&&\text{\axiomref{def:causal_space}(ii)}\\
            &=\lim_{n\to\infty}\sum^{m_n}_{i_n=1}b_{i_n}\int\iPP{U}{\QQ}(d\omega)\ind_B(\omega)\ind_H(\omega)\ind_{H_{i_n}}(\omega)\\
            &=\int\iPP{U}{\QQ}(d\omega)\ind_B(\omega)\ind_H(\omega)\lim_{n\to\infty}\sum^{m_n}_{i_n=1}b_{i_n}\ind_{H_{i_n}}(\omega)&&\text{monotone convergence}\\
            &=\int\iPP{U}{\QQ}(d\omega)\ind_B(\omega)\ind_H(\omega)\PP_\sG(\omega,A)\\
            &=\iEE{U}{\QQ}\left[\ind_B\ind_H\PP_\sG(\cdot,A)\right]\\
            &=\iEE{U}{\QQ}\left[\ind_H\iPP{U}{\QQ}_\sG(\cdot,B)\PP_\sG(\cdot,A)\right],
        \end{alignat*}
        where the last line follows from the definition of the conditional probability measure \(\iPP{U}{\QQ}_\sG(\cdot,B)\), since the map \(\omega'\mapsto\ind_H(\omega')\PP_\sG(\omega',A)\) is \(\sG\)-measurable. But we saw earlier that for \(\iPP{U}{\QQ}\)-almost all \(\omega'\in\Omega\), we have \(\iPP{U}{\QQ}_\sG(\omega',A)=\PP_\sG(\omega',A)\). Hence, for all \(H\in\sG\), we have
        \[\iEE{U}{\QQ}\left[\ind_H\iPP{U}{\QQ}_\sG(\cdot,A\cap B)\right]=\iEE{U}{\QQ}\left[\ind_H\iPP{U}{\QQ}_\sG(\cdot,B)\iPP{U}{\QQ}_\sG(\cdot,A)\right].\]
        Hence, for \(\iPP{U}{\QQ}\)-almost all \(\omega'\in\Omega\), we have
        \[\iPP{U}{\QQ}_\sG(\omega',A\cap B)=\iPP{U}{\QQ}_\sG(\omega',B)\iPP{U}{\QQ}_\sG(\omega',A),\]
        in other words, \(A\) and \(B\) are conditionally independent given \(\sG\) under the measure \(\iPP{U}{\QQ}\). Since \(B\in\sH_U\) was arbitrary, we have that \(\sH_U\) and \(A\) are conditionally independent given \(\sG\) under \(\iPP{U}{\QQ}\), as required. 
    \end{enumerate}
\end{proof}

\printprop{prop:post-intervention}
\begin{proof}
    By the definition of active post-intervention causal effects, we have, for all \(\omega_{V\setminus U}\in\Omega_{V\setminus U}\),
    \[K_{U\cup V}(\omega_{U\cup V},A)=K_V(\omega_V,A).\tag{*}\]
    We will use this property in both parts of the proof. 
    \begin{enumerate}[(i)]
        \item Since \(U\cap V=\emptyset\), we have \(V\setminus U=V\). Then see that, for all \(\omega_U\in\Omega_U\),
        \begin{alignat*}{3}
            \iK{V}{\QQ}{U}(\omega_U,A)&=\int\QQ(d\omega'_{V\setminus U})K_{U\cup V}((\omega_U,\omega'_{V\setminus U}),A)\quad&&\text{\definitionref{def:interventions}}\\
            &=\int\QQ(d\omega'_V)K_V(\omega'_V,A)&&\text{by (*)}\\
            &=\iPP{V}{\QQ}(A)&&\text{\definitionref{def:interventions}}.
        \end{alignat*}
        This is precisely the definition of no active causal effect in \definitionref{def:active_causal_effect}. 
        \item Since \(U\subseteq U\cup V\), we have \(U\setminus(U\cup V)=\emptyset\). Hence,
        \[\iK{U}{\QQ}{U\cup V}(\omega,A)=\int\QQ(d\omega'_\emptyset)K_{U\cup V}((\omega_{U\cup V},\omega_\emptyset),A)=K_{U\cup V}(\omega,A).\]
        But then by (*), we have
        \[\iK{U}{\QQ}{U\cup V}(\omega,A)=K_V(\omega,A).\]
        On the other hand, we have
        \begin{alignat*}{3}
            \iK{U}{\QQ}{V}(\omega,A)&=\int\QQ(d\omega'_{U\setminus V})K_{U\cup V}((\omega_V,\omega'_{U\setminus V}),A)\\
            &=\int\QQ(d\omega'_{U\setminus V})K_V(\omega,A)&&\text{by (*)}\\
            &=K_V(\omega,A).
        \end{alignat*}
        Putting the two together, we have
        \[\iK{U}{\QQ}{U\cup V}(\omega,A)=K_V(\omega,A).\]
        This is precisely the definition of no active post-intervention causal effect in \definitionref{def:active_causal_effect_post-intervention}. 
    \end{enumerate}
\end{proof}

\end{document}